\RequirePackage{lhchiggs}
\documentclass[letterpaper]{cernyrep}
\usepackage{rotating}
\usepackage{color}
\usepackage{subfigure}
\usepackage{lscape}
\usepackage{floatpag}
\floatpagestyle{plain}
\usepackage{axodraw}
\usepackage{lineno}
\newcommand{\hcdot}{\ensuremath{\hfill}}
\newcommand{\tc}[2]{\textcolor{#1}{#2}}
\newcommand{\BRinv}{\ensuremath{\mathrm{BR_{inv.,undet.}}}\xspace}
\newcommand{\Cc}{\ensuremath{\upkappa}}
\newcommand{\Rr}{\ensuremath{\uplambda}}

\newcommand{\MyggH}{\ensuremath{\Pg\Pg\PH}}
\newcommand{\MyttH}{\ensuremath{\PQt\PAQt\PH}}

\DeclareRobustCommand{\PV}{\HepParticle{V}{}{}\Xspace} % vector

\DeclareRobustCommand{\Pf}{\HepParticle{f}{}{}\Xspace} % fermion (generic)
\DeclareRobustCommand{\PAf}{\HepAntiParticle{\Pf}{}{}\Xspace} % anti-fermion (generic)

\newcommand{\mH}{\ensuremath{m_{\PH}}}

\newcommand{\HIGLU}{{\sl HIGLU\xspace}}
\newcommand{\HDECAY}{{\sl HDECAY\xspace}}
\newcommand{\GGHNNLO}{{\sl GGH@NNLO\xspace}}

\newcommand{\bb}{\ensuremath{\PQb\PAQb}}
\newcommand{\ggF}{\ensuremath{\Pg\Pg\to \PH}}
\newcommand{\VBF}{\ensuremath{\PQq\PQq^\prime\to \PQq\PQq^\prime\PH}}
\newcommand{\VH}{\ensuremath{\PQq\PAQq\to \PW\PH/\PZ\PH}}
\newcommand{\ttH}{\ensuremath{\PQq\PAQq/\Pg\Pg \to \PQt\PAQt\PH}}

\newcommand{\hgg}{\ensuremath{\PH \to \PGg\PGg}}

\newcommand{\hWWlnln}{\ensuremath{\PH \to \PW\PW^{(*)} \to
    \ell^+\nu\ell^-\overline{\nu}}}

 \newcommand{\hZZllll}{\ensuremath{\PH
    \to \PZ\PZ^{(*)}\to \ell^+\ell^-\ell^+\ell^-}}
\newcommand{\htt}{\ensuremath{\PH \to \tau^+\tau^-}}
\newcommand{\hbb}{\ensuremath{\PH \to \bb}}

\begin{document}

% --- line numbers and draft heading
%\linenumbers
%\draftheading
  
% --- print draft version
\begin{flushright}
   LHCHXSWG-2012-001 \\
\end{flushright}

\pagenumbering{roman}

\title{LHC HXSWG interim recommendations to explore the coupling structure of a Higgs-like particle}

\author{LHC Higgs Cross Section Working Group, Light Mass Higgs Subgroup\\
A.~David$^{\,1}$, A.~Denner$^{\,2}$, M.~D\"uhrssen$^{\,3}$, M.~Grazzini$^{\,4}$\footnote{On leave of absence from INFN, Sezione di Firenze, Sesto Fiorentino, Florence, Italy}, C.~Grojean$^{\,3}$, G.~Passarino$^{\,5}$, M.~Schumacher$^{\,6}$, M.~Spira$^{\,7}$, G.~Weiglein$^{\,8}$, and M.~Zanetti$^{\,9}$.}

\date{August 2012}

\maketitle % this produces the title block

\begin{itemize}

\item[$^{1}$] Laborat\'orio de Instrumentac\~ao e F\'\i sica Experimental de Part\'\i culas, Lisboa, Portugal

\item[$^{2}$] Universit\"at W\"urzburg, Institut f\"ur Theoretische Physik und Astrophysik,  W\"urzburg, Germany

\item[$^{3}$] CERN, Geneva, Switzerland

\item[$^{4}$] Institut f\"ur Theoretische Physik, Universit\"at Z\"urich, Z\"urich, Switzerland

\item[$^{5}$] Dipartimento di Fisica Teorica, Universit\`a di Torino and INFN, Sezione di Torino, Torino, Italy

\item[$^{6}$] Fakult\"at f\"ur Mathematik und Physik, Albert-Ludwigs-Universit\"at, Freiburg, Germany

\item[$^{7}$] Paul Scherrer Institut, W\"urenling und Villigen, Villigen PSI, Switzerland

\item[$^{8}$] DESY, Hamburg, Germany

\item[$^{9}$] Massachusetts Institute of Technology, Cambridge, USA

\end{itemize}

\vskip 0.5cm
\begin{abstract}
This document presents an interim framework in which the coupling structure of a Higgs-like particle can be studied.
After discussing different options and approximations, recommendations
on specific benchmark parametrizations to be used to fit the data are given.
% Future directions on this topic are briefly discussed.
\end{abstract}

% --- Introduction
\newpage
\pagenumbering{arabic}
\setcounter{footnote}{0}
%
% --- Higgs subsections

\tableofcontents

%\newpage
\section{Introduction}

The recent observation of a new massive neutral boson by ATLAS and CMS~\cite{HiggsObsATLAS, HiggsObsCMS},
as well as
evidence from the Tevatron experiments~\cite{HiggsObsTevatron}, opens a new era where characterization
of this new object is of central importance.

The Standard Model (SM), as any renormalizable theory,
makes very accurate predictions for the coupling of the Higgs boson
to all other known particles.
These couplings directly influence the rates of production and decay of the Higgs boson.
Therefore, measurement of the production and decay rates of the observed state yields information
that can be used to probe whether data are compatible with the SM predictions for the Higgs boson.

While coarse features of the observed state can be inferred from the information that the experiments have made public,
only a consistent and combined treatment of the data can yield the most accurate picture of the coupling structure.
Such a treatment must take into account all the systematic and statistical uncertainties considered in the analyses,
as well as the correlations among them. 

This document outlines an interim framework to explore the coupling structure of the recently
observed state.
The framework proposed in this recommendation should be seen as a continuation of the earlier studies of the LHC sensitivity to the Higgs couplings initiated in Refs.~\cite{Zeppenfeld:2000td, Duhrssen:2003aa, Duhrssen:2004cv, Lafaye:2009vr}, and has been influenced by the works of Refs.~\cite{Hagiwara:1993qt, GonzalezGarcia:1999fq, Barger:2003rs, Manohar:2006gz, Giudice:2007fh, Low:2009di, Espinosa:2010vn, Anastasiou:2011pi}.  It follows closely the methodology proposed in the recent phenomenological works of Refs.~\cite{Carmi:2012yp, Azatov:2012bz, Espinosa:2012ir} which have been further extended in several directions \cite{Cao:2012fz, Boudjema:2012cq, Barger:2012hv, Frandsen:2012rj, Giardino:2012ww, Ellis:2012rx, Draper:2012xt, Azatov:2012rd, Farina:2012ea, Englert:2012cb, Degrande:2012gr, Klute:2012pu, Arhrib:2012yv, Akeroyd:2012ms, Azatov:2012wq, Carena:2012xa, Gupta:2012mi, Blum:2012ii, Chang:2012tb, Chang:2012gn, Low:2012rj, Benbrik:2012rm, Corbett:2012dm, Giardino:2012dp, Ellis:2012hz, Montull:2012ik, Espinosa:2012im, Carmi:2012in, Peskin:2012we, Banerjee:2012xc, Cao:2012yn, Bertolini:2012gu, ArkaniHamed:2012kq, Bonnet:2012nm, Craig:2012vn, Almeida:2012bq, Alves:2012ez, Espinosa:2012in, Accomando:2012yg, Elander:2012fk, Reece:2012gi, Djouadi:2012rh} along the lines that are formalized in the present recommendation.
While the interim framework is not final,
it has an accuracy that matches the statistical power of the datasets that the LHC experiments
can hope to collect until the end of the 2012 LHC run
and is an explicit attempt to provide a common ground
for the dialogue in the, and between the, experimental and theoretical communities. 

Based on that framework, a series of benchmark parametrizations are presented.
Each benchmark parametrization allows to explore specific aspects of the coupling structure of the new state.
The parametrizations have varying degrees of complexity,
with the aim to cover the most interesting possibilities that can be realistically tested with
the LHC 7 and 8\UTeV\ datasets.
On the one hand, the framework and benchmarks were designed to provide a recommendation
to experiments on how to perform coupling fits that are useful for the theory community.
On the other hand the theory community can prepare for results based on the framework discussed in this document.

%The recommended benchmarks in this document were designed to provide a solid basis on which two aspects can be developed.
%On the one hand, the recommendation can be used as basis by the experiments to combine their results according to clearly defined parametrizations.
%On the other hand, theoretical work can be developed with knowledge of which parametrizations experiments will
%use to publish their results. 

%Finally, avenues that can be followed to improve upon this interim framework are briefly expounded upon.
Finally, avenues that can be pursued to improve upon this interim framework and
recommendations on how to probe the tensor structure will be discussed in a future document.

\section{Panorama of experimental measurements at the LHC}

In 2011, the LHC delivered an integrated luminosity of slightly less
than 6\ifb\ of proton--proton ($\Pp\Pp$) collisions at a center-of-mass
energy of 7\UTeV\ to the ATLAS and CMS experiments. 
By July 2012, the LHC delivered more than 6\ifb\ of $\Pp\Pp$ collisions
at a center-of-mass energy of 8\UTeV\ to both experiments. 
For this dataset, the instantaneous luminosity reached
record levels of approximately $7 \cdot 10^{33}$~cm$^{-2}$s$^{-1}$,
almost double the peak luminosity of 2011 with the same 50~ns bunch
spacing.
The 2012 $\Pp\Pp$ run will continue until the end of
the year, hopefully delivering about 30\ifb\ per
experiment.
 
At the LHC a SM-like Higgs boson is searched for mainly in four
exclusive production processes:
the predominant gluon fusion \ggF,
the vector boson fusion \VBF,
the associated production with a vector boson \VH\ and
the associated production with a top-quark pair \ttH.
The main search
channels are determined by five decay modes of the Higgs boson, the
$\PGg\PGg$, $\PZ\PZ^{(*)}$, $\PW\PW^{(*)}$, $\PQb\PAQb$ and
$\PGtp\PGtm$ channels.
The mass range within which each channel is effective
and the production processes for which exclusive searches
have been developed and made public
are indicated in Table~\ref{tab:channels}.
A detailed description of the Higgs search analyses
can be found in~\Brefs{HiggsObsATLAS,HiggsObsCMS}. 

\begin{table}[htb]
\centering
  \caption{Summary of the Higgs boson search channels in
    the ATLAS and CMS experiments by July 2012.
    The $\surd$ symbol indicates exclusive searches targetting the inclusive \ggF\ production, 
    the associated production processes (with a vector boson or a top quark pair)
    or the vector boson fusion (VBF) production process.}
\label{tab:channels}
\begin{tabular}{lc|cc|cc|cc|cc}\hline
  Channel & $\mH (\UGeV)$ &  \multicolumn{2}{c}{\MyggH} &
  \multicolumn{2}{c}{VBF} & \multicolumn{2}{c}{VH} &
  \multicolumn{2}{c}{\MyttH} \\ \hline
%  & & \tiny ATLAS & \tiny CMS  & \tiny ATLAS & \tiny CMS  & \tiny ATLAS & \tiny CMS  & \tiny ATLAS & \tiny CMS   \\ \hline
  & & \scriptsize ATLAS & \scriptsize CMS  & \scriptsize ATLAS & \scriptsize CMS  & \scriptsize ATLAS & \scriptsize CMS  & \scriptsize ATLAS & \scriptsize CMS   \\ \hline
  \hgg\     & 110--150 & $\surd$ & $\surd$ & $\surd$ & $\surd$  & - & -  & - & -       \\ \hline
  \htt\     & 110--145 & $\surd$ & $\surd$ & $\surd$ & $\surd$  & $\surd$ & $\surd$ & - & -       \\ \hline
  \hbb\     & 110--130 & - & - & - & -  & $\surd$ & $\surd$ & - & $\surd$ \\ \hline
  \hZZllll\ & 110--600 & $\surd$ & $\surd$ & - & -  & - & -  & - & -       \\ \hline
  \hWWlnln\ & 110--600 & $\surd$ & $\surd$ & $\surd$ & $\surd$  & $\surd$ & $\surd$ & - & -       \\ \hline
\end{tabular}
\end{table}

Both the ATLAS and CMS experiments observe an excess of events for
Higgs boson mass hypotheses near $\sim 125\UGeV$.
The observed combined significances are $5.9\sigma$ for ATLAS~\cite{HiggsObsATLAS}
and 5.0$\sigma$ for CMS~\cite{HiggsObsCMS}, compatible with their respective sensitivities.
Both observations are primarily in the \hgg, \hZZllll\ and \hWWlnln\ channels.
For the \hgg\ channel, excesses of $4.5\sigma$ and $4.1\sigma$ are observed
at Higgs boson mass hypotheses of $126.5\UGeV$ and $125\UGeV$,
in agreement with the expected sensitivities of around $2.5\sigma$ and $2.8\sigma$,
in the ATLAS and CMS experiments respectively.
For the \hZZllll\ channel, the significances of the excesses are 3.6$\sigma$ and 3.2$\sigma$
at Higgs boson mass hypotheses of $125\UGeV$ and $125.6\UGeV$,
in the ATLAS and CMS experiments respectively.
The expected sensitivities at those masses are 2.7$\sigma$ in ATLAS and 3.8$\sigma$
in CMS respectively.
For the low mass resolution \hWWlnln\ channel ATLAS observes an excess of 2.8$\sigma$ (2.3$\sigma$ expected) 
and CMS observes 1.6$\sigma$ (2.4$\sigma$ expected) for a Higgs boson mass hypotheses of $\sim 125\UGeV$.
The other channels do not contribute
significantly to the excess, but are nevertheless individually
compatible with the presence of a signal.

The ATLAS and CMS experiments have also reported compatible
measurements of the mass of the observed narrow resonance yielding:
\begin{center}
  126.0~$\pm 0.4$(stat.)~$\pm 0.4$(syst.)\UGeV (ATLAS), \\
  125.3~$\pm 0.4$(stat.)~$\pm 0.5$(syst.)\UGeV (CMS).
\end{center}

\section{Interim framework for the search of deviations}
The idea behind this framework is that all deviations from the SM
are computed assuming that there is only one underlying state at $\sim125\UGeV$.
It is assumed that this state is a Higgs boson,
i.e.\ the excitation of a field whose vacuum expectation value (VEV) breaks electroweak symmetry,
and that it is SM-like,
in the sense that the experimental results so far are compatible
with the interpretation of the state in terms of the SM Higgs boson.
No specific assumptions are made on any additional states of new physics
(and their decoupling properties)
that could influence the phenomenology of the $125\UGeV$ state,
such as additional Higgs bosons (which could be heavier but also lighter than $125\UGeV$),
additional scalars that do not develop a VEV,
and new fermions and/or gauge bosons that could interact with the state at $125\UGeV$, 
giving rise, for instance, to an invisible decay mode.

The purpose of this framework is to
either confirm that the light, narrow, resonance indeed matches the properties of the SM Higgs,
or to establish a deviation from the SM behaviour, which would rule out the SM if sufficiently
significant.
In the latter case the next goal in the quest to identify the nature of
electroweak symmetry breaking (EWSB) would obviously
be to test the compatibility of the observed patterns with alternative frameworks of EWSB. 

% The idea behind our framework is that all deviations from SM are computed assuming that there 
% is only one underlying state at $\sim125\UGeV$ which we assume to be a SM-like Higgs boson,
% i.e.\ the excitation of a field whose vacuum expectation value (VEV) breaks electroweak symmetry. 
% We make no specific assumption on the existence and nature of other heavy (scalar or not) degrees 
% of freedom which can influence the SM-like Higgs boson couplings to all SM particles; furthermore
% no assumption is made on their decoupling as their masses increase or on the mass-mixing with
% the SM-like Higgs boson. 
% 
% The heavy scalar degrees of freedom are Higgs-partners: they are not in the SM, but may
% have a rich spectrum of non-Higgs states (they do not necessarily develop a VEV). 
% Their spectrum and couplings to fermions and vector bosons will be strongly model dependent and
% our frameworks are intended to be part of a larger program (the so-called inverse problem):
% if LHC finds evidence for physics beyond the SM, how can one determine the underlying theory?
% Therefore, our framework is designed for proving that the light, narrow, resonance matches 
% the SM Higgs properties, or to establish that deviations from the SM behaviour are consistent 
% with some other EWSB framework.

In investigating the experimental information that can be obtained on the coupling properties
of the new state near $125\UGeV$ from the LHC data to be collected in 2012
the following assumptions are made\footnote{The experiments are encouraged to test the assumptions of the framework, but that lies outside the scope of this document.}:
\begin{itemize}
 \item The signals observed in the different search channels
originate from a single narrow resonance with a mass near $125\UGeV$. 
The case of several, possibly overlapping, resonances in this mass 
region is not considered.
 \item The width of the assumed Higgs boson near $125\UGeV$ is neglected, 
i.e.\ the zero-width approximation for this state is used.
Hence the
%product $\sigma\times BR(\mathit{ii}\to\PH\to\mathit{ff})$
signal cross section
can be decomposed in the following way for all channels:
 \begin{equation}
\left(\sigma\cdot\text{BR}\right)(\mathit{ii}\to\PH\to\mathit{ff}) = \frac{\sigma_{\mathit{ii}}\cdot\Gamma_{\mathit{ff}}}{\Gamma_{\PH}}
 \end{equation}
 where $\sigma_{\mathit{ii}}$ is the production cross section through the initial state $\mathit{ii}$, $\Gamma_{\mathit{ff}}$ the partial decay width into the final state $\mathit{ff}$ and $\Gamma_{\PH}$ the total width of the Higgs boson.
\end{itemize}

Within the context of these assumptions, in the following
a simplified framework for investigating the
experimental information that can be obtained on the coupling properties
of the new state is outlined.
In general, the couplings of the assumed Higgs state near $125\UGeV$ are
``pseudo-observables'', i.e.\ they cannot be directly measured. 
This means that a certain ``unfolding procedure'' is
necessary to extract information on the couplings from the
measured quantities like cross sections times branching ratios (for
specific experimental cuts and acceptances).
This gives rise to a certain model dependence of the extracted information.
Different options can be pursued in this context.
One possibility is to confront a specific model with the experimental data.
This has the advantage that all available higher-order corrections within this model can
consistently be included and also other experimental
constraints (for instance from direct searches or from electroweak
precision data) can be taken into account.
However, the results obtained in this case are restricted to the interpretation within
that particular model.
Another possibility is to use a general
parametrization of the couplings of the new state without referring to any
particular model.
While this approach is clearly less model-dependent,
the relation between the extracted coupling parameters and the couplings of actual models, 
for instance the SM or its minimal
supersymmetric extension (MSSM), is in general non-trivial, so that the
theoretical interpretation of the extracted information can
be difficult. 
It should be mentioned that the results for the 
signal strengths of individual search channels that have been made
public by ATLAS and CMS, while referring just to a particular search
channel rather than to the full information available from the Higgs
searches, are nevertheless very valuable for testing the predictions
of possible models of physics beyond the SM.

In the SM, once the numerical value of the Higgs
mass is specified, all the couplings of the Higgs boson to fermions,
bosons and to itself are specified within the model.
It is therefore in
general not possible to perform a fit to experimental data within the
context of the SM where Higgs couplings are treated as free parameters. 
While it is possible to test the overall compatibility of the SM with
the data, it is not possible to extract information about 
deviations of the measured couplings with respect to their SM values. 

A theoretically well-defined framework for probing small deviations from the
SM predictions --- or the predictions of another reference model --- is to use
the state-of-the-art predictions in this model (including all available
higher-order corrections) and to supplement them with the contributions
of additional terms in the Lagrangian,
which are usually called ``anomalous couplings''.
In such an approach and in general, not only the coupling strength, i.e.\ the
absolute value of a given coupling, will be modified, but also the
tensor structure of the coupling.
For instance, the $\PH\PWp\PWm$ LO coupling 
in the SM is proportional to the metric tensor $g^{\mu\nu}$, while
anomalous couplings will generally also give rise to other tensor
structures, however required to be compatible with the SU(2)$\times$U(1)
symmetry and the corresponding Ward-Slavnov-Taylor identities.
As a consequence, kinematic distributions will in general
be modified when compared to the SM case. 

Since the reinterpretation of searches that have been performed within
the context of the SM is difficult if effects that change kinematic
distributions are taken into account and since not all the necessary
tools to perform this kind of analysis are available yet, the
following additional assumption is made in this simplified framework:

\begin{itemize}

\item
Only modifications of couplings strengths, i.e.\ of absolute values of
couplings, are taken into account, while the tensor structure of the
couplings is assumed to be the same as in the SM prediction. This means in
particular that the observed state is assumed to be a CP-even scalar.

\end{itemize}

\subsection{Definition of coupling scale factors}
\label{sec:scale_factor_def}

In order to take into account the currently best available SM
predictions for Higgs cross sections, which include
higher-order QCD and EW corrections~\cite{Dittmaier:2011ti,Dittmaier:2012vm,HiggsWG},
while at the same time introducing
possible deviations from the SM values of the couplings, the 
predicted SM Higgs cross sections and partial
decay widths are dressed with scale factors $\Cc_i$. 
The scale factors $\Cc_i$ are defined in such a way that the cross sections
$\sigma_{ii}$ or the partial decay widths $\Gamma_{ii}$ associated with the SM particle $i$ scale
with the factor $\Cc_i^2$ when compared to the corresponding SM prediction.
Table~\ref{tab:LO_coupling_relatios} lists all relevant cases.
Taking the process $\Pg\Pg\to\PH\to\PGg\PGg$ as an example, one would use as cross section:
\begin{eqnarray}
\left(\sigma\cdot\text{BR}\right)(\Pg\Pg\to\PH\to\PGg\PGg) &=& \sigma_\text{SM}(\Pg\Pg\to\PH) \cdot \text{BR}_\text{SM}(\PH\to\PGg\PGg)\,\cdot \frac{\Cc_{\Pg}^2 \cdot \Cc_{\PGg}^2}{\Cc_{\PH}^2}
\end{eqnarray}
where the values and uncertainties for both $\sigma_\text{SM}(\Pg\Pg\to\PH)$
and $\text{BR}_\text{SM}(\PH\to\PGg\PGg)$ are taken from Ref.~\cite{HiggsWG} for a given Higgs mass hypothesis.

\begin{table}
\centering
\begin{tabular}{@{}p{0.43\linewidth}|p{0.55\linewidth}@{}}
\hline
\begin{eqnarray}
  \omit\rlap{\text{Production modes}}\nonumber\\
  \frac{\sigma_{\MyggH}}{\sigma_{\MyggH}^\text{SM}}               & = & \left\{ \begin{array}{l} \Cc_{\Pg}^2(\Cc_{\PQb},\Cc_{\PQt},\mH) \\ \Cc_{\Pg}^2 \end{array} \right. \\
  \frac{\sigma_{\text{VBF}}}{\sigma_{\text{VBF}}^\text{SM}} & = & \Cc_\text{VBF}^2(\Cc_{\PW},\Cc_{\PZ},\mH)\\
  \frac{\sigma_{\PW\PH}}{\sigma_{\PW\PH}^\text{SM}}                 & = & \Cc_{\PW}^2\\
  \frac{\sigma_{\PZ\PH}}{\sigma_{\PZ\PH}^\text{SM}}                 & = & \Cc_{\PZ}^2\\
  \frac{\sigma_{\MyttH}}{\sigma_{\MyttH}^\text{SM}}    & = & \Cc_{\PQt}^2
\end{eqnarray}
&
% \begin{minipage}[t]{\linewidth}
% Detectable decay modes
\begin{eqnarray}
%   \hline
  \omit\rlap{\text{Detectable decay modes}}\nonumber\\
  \frac{\Gamma_{\PW\PW^{(*)}}}{\Gamma_{\PW\PW^{(*)}}^\text{SM}}           &=& \Cc_{\PW}^2\\
  \frac{\Gamma_{\PZ\PZ^{(*)}}}{\Gamma_{\PZ\PZ^{(*)}}^\text{SM}}           &=& \Cc_{\PZ}^2\\
  \frac{\Gamma_{\PQb\PAQb}}{\Gamma_{\PQb\PAQb}^\text{SM}}           &=& \Cc_{\PQb}^2\\
  \frac{\Gamma_{\PGtm\PGtp}}{\Gamma_{\PGtm\PGtp}^\text{SM}}     &=& \Cc_{\PGt}^2\\
  \frac{\Gamma_{\PGg\PGg}}{\Gamma_{\PGg\PGg}^\text{SM}} &=& \left\{ \begin{array}{l} \Cc_{\PGg}^2(\Cc_{\PQb},\Cc_{\PQt},\Cc_{\PGt},\Cc_{\PW},\mH) \\ \Cc_{\PGg}^2 \end{array} \right.\\
  \frac{\Gamma_{\PZ\PGg}}{\Gamma_{\PZ\PGg}^\text{SM}} &=& \left\{ \begin{array}{l} \Cc_{(\PZ\PGg)}^2(\Cc_{\PQb},\Cc_{\PQt},\Cc_{\PGt},\Cc_{\PW},\mH) \\ \Cc_{(\PZ\PGg)}^2 \end{array} \right.\\
% \end{eqnarray}
% 
% Undetectable decay modes
% \begin{eqnarray}  
%   \hline
  \nonumber\\
  \omit\rlap{\text{Currently undetectable decay modes}}\nonumber\\
  \frac{\Gamma_{\PQt\PAQt}}{\Gamma_{\PQt\PAQt}^\text{SM}}           &=& \Cc_{\PQt}^2\\
  \frac{\Gamma_{\Pg\Pg}}{\Gamma_{\Pg\Pg}^\text{SM}}           &:& \text{see Section~\ref{sec:C_g}}\nonumber\\
  \frac{\Gamma_{\PQc\PAQc}}{\Gamma_{\PQc\PAQc}^\text{SM}}           &=& \Cc_{\PQt}^2\\
  \frac{\Gamma_{\PQs\PAQs}}{\Gamma_{\PQs\PAQs}^\text{SM}}           &=& \Cc_{\PQb}^2\\
  \frac{\Gamma_{\PGmm\PGmp}}{\Gamma_{\PGmm\PGmp}^\text{SM}}       &=& \Cc_{\PGt}^2\\
% \end{eqnarray}
% 
% Total width
% \begin{eqnarray}    
%   \hline
  \nonumber\\
  \omit\rlap{\text{Total width}}\nonumber\\
  \frac{\Gamma_{\PH}}{\Gamma_{\PH}^\text{SM}}           &=& \left\{ \begin{array}{l} \Cc_{\PH}^2(\Cc_i,\mH) \\ \Cc_{\PH}^2 \end{array} \right.                    
\end{eqnarray}
% \end{minipage}
\\
\hline
\end{tabular}
\caption{LO coupling scale factor relations for Higgs boson cross sections and partial decay widths relative to the SM.
For a given $\mH$ hypothesis, the smallest set of degrees of freedom in this framework comprises
$\Cc_{\PW}$, $\Cc_{\PZ}$, $\Cc_{\PQb}$, $\Cc_{\PQt}$, and $\Cc_{\PGt}$.
For partial widths that are not detectable at the LHC, scaling is performed via proxies chosen among the detectable ones.  
Additionally, the loop-induced vertices can be treated as a function of other $\Cc_i$ or effectively,
through the $\Cc_{\Pg}$ and $\Cc_{\PGg}$ degrees of freedom which allow probing for BSM contributions in the loops.
Finally, to explore invisible or undetectable decays,
the scaling of the total width can also be taken as a separate degree of freedom, $\Cc_{\PH}$,
instead of being rescaled as a function, $\Cc_{\PH}^2(\Cc_i,\mH)$, of the other scale factors.}
\label{tab:LO_coupling_relatios}
\end{table}

By definition, the currently best available SM predictions
for all $\sigma\cdot\text{BR}$ are recovered when all $\Cc_i=1$.
In general, this means that for $\Cc_i\neq 1$ higher-order accuracy is lost.
Nonetheless, NLO QCD corrections essentially factorize with respect to coupling rescaling,
and are accounted for wherever possible.
This approach ensures that for a true SM Higgs boson no artifical deviations
(caused by ignored NLO corrections) are found from what is considered the SM Higgs boson hypothesis.
The functions
$\Cc_\text{VBF}^2(\Cc_{\PW},\Cc_{\PZ},\mH)$,
$\Cc_{\Pg}^2(\Cc_{\PQb},\Cc_{\PQt},\mH)$,
$\Cc_{\PGg}^2(\Cc_{\PQb},\Cc_{\PQt},\Cc_{\PGt},\Cc_{\PW},\mH)$ and
$\Cc_{\PH}^2(\Cc_i,\mH)$
 are used for cases where there is a non-trivial relationship between scale factors $\Cc_i$ and cross sections
 or (partial) decay widths, and are calculated to NLO QCD accuracy.
 The functions are defined in the following sections and 
 all required input parameters as well as example code can be found in~\Brefs{HiggsWG,LMwiki}.
As explained in Sec.~\ref{Subsub:int} below, the
notation in terms of the partial widths $\Gamma_{\PW\PW^{(*)}}$ and $\Gamma_{\PZ\PZ^{(*)}}$ in Table~\ref{tab:LO_coupling_relatios}
is meant for illustration only. In the experimental analysis the
4-fermion partial decay widths are taken into account.

\subsubsection{Scaling of the VBF cross section}
$\Cc_\text{VBF}^2$ refers to the functional dependence of the
VBF\footnote{Vector Boson Fusion is also called Weak Boson Fusion, as only the weak bosons $\PW$ and $\PZ$ contribute to the production.}
cross section on the scale factors $\Cc_{\PW}^2$ and $\Cc_{\PZ}^2$:
\begin{eqnarray}
\label{eq:RVBF}
 \Cc_\text{VBF}^2(\Cc_{\PW},\Cc_{\PZ},\mH) &=& \frac{\Cc_{\PW}^2 \cdot \sigma_{\PW F}(\mH) + \Cc_{\PZ}^2 \cdot \sigma_{\PZ F}(\mH)}{\sigma_{\PW F}(\mH)+\sigma_{\PZ F}(\mH)}
\end{eqnarray}
The $\PW$- and $\PZ$-fusion cross sections,
$\sigma_{\PW F}$ and $\sigma_{\PZ F}$,
are taken from \Brefs{Arnold:2008rz,webVBFNLO}.
The interference term is $< 0.1\%$ in the SM and hence ignored~\cite{Ciccolini:2007ec}.

\subsubsection{Scaling of the gluon fusion cross section and of the $\PH\to\Pg\Pg$ decay vertex}
\label{sec:C_g}
$\Cc_{\Pg}^2$ refers to the scale factor for the loop-induced production
cross section $\sigma_{\MyggH}$.
The decay width $\Gamma_{\Pg\Pg}$ is not observable at the LHC, however its contribution to the total width
is also considered.

\paragraph*{Gluon fusion cross-section scaling}
As NLO QCD corrections factorize with the scaling of the electroweak couplings with $\Cc_{\PQt}$ and $\Cc_{\PQb}$,
the function $\Cc_{\Pg}^2(\Cc_{\PQb}, \Cc_{\PQt},\mH)$ can be calculated in NLO QCD:
\begin{eqnarray}
\label{eq:CgNLOQCD}
 \Cc_{\Pg}^2(\Cc_{\PQb}, \Cc_{\PQt},\mH) &=& \frac{\Cc_{\PQt}^2\cdot\sigma_{\MyggH}^{\PQt\PQt}(\mH) +\Cc_{\PQb}^2\cdot\sigma_{\MyggH}^{\PQb\PQb}(\mH) +\Cc_{\PQt}\Cc_{\PQb}\cdot\sigma_{\MyggH}^{\PQt\PQb}(\mH)}{\sigma_{\MyggH}^{\PQt\PQt}(\mH)+\sigma_{\MyggH}^{\PQb\PQb}(\mH)+\sigma_{\MyggH}^{\PQt\PQb}(\mH)}
\end{eqnarray}

Here, $\sigma_{\MyggH}^{\PQt\PQt}$, $\sigma_{\MyggH}^{\PQb\PQb}$ and $\sigma_{\MyggH}^{\PQt\PQb}$
denote the square of the top-quark contribution, the square of the bottom-quark
contribution and the top-bottom interference, respectively.
The interference term ($\sigma_{\MyggH}^{\PQt\PQb}$) is negative for a light mass Higgs, $\mH < 200\UGeV$. 
Within the LHC Higgs Cross Section Working Group (for the evaluation of the
MSSM cross section) these contributions were evaluated, where for
$\sigma_{\MyggH}^{\PQb\PQb}$ and $\sigma_{\MyggH}^{\PQt\PQb}$ the full NLO QCD calculation included
in \HIGLU~\cite{Spira:1995mt} was used.
For $\sigma_{\MyggH}^{\PQt\PQt}$ the NLO QCD result of \HIGLU\ was supplemented with the
NNLO corrections in the heavy-top-quark limit as implemented in \GGHNNLO~\cite{Harlander:2002wh},
see~Ref.~\cite[Sec.~6.3]{Dittmaier:2011ti} for details.

\paragraph*{Partial width scaling}
In a similar way, NLO QCD corrections for the
$\PH\to \Pg\Pg$ partial width are implemented in \HDECAY~\cite{Spira:1996if,Djouadi:1997yw,hdecay2}.
This allows to treat the scale factor for $\Gamma_{\Pg\Pg}$
as a second order polynomial in $\Cc_{\PQb}$ and $\Cc_{\PQt}$:
\begin{eqnarray}
\label{eq:GammagNLOQCD}
\frac{\Gamma_{\Pg\Pg}}{\Gamma_{\Pg\Pg}^\text{SM}(\mH)} = \frac{\Cc_{\PQt}^2\cdot\Gamma_{\Pg\Pg}^{\PQt\PQt}(\mH) +\Cc_{\PQb}^2\cdot\Gamma_{\Pg\Pg}^{\PQb\PQb}(\mH) +\Cc_{\PQt}\Cc_{\PQb}\cdot\Gamma_{\Pg\Pg}^{\PQt\PQb}(\mH)}{\Gamma_{\Pg\Pg}^{\PQt\PQt}(\mH)+\Gamma_{\Pg\Pg}^{\PQb\PQb}(\mH)+\Gamma_{\Pg\Pg}^{\PQt\PQb}(\mH)}
\end{eqnarray}
The terms $\Gamma_{\Pg\Pg}^{\PQt\PQt}$, $\Gamma_{\Pg\Pg}^{\PQb\PQb}$ and $\Gamma_{\Pg\Pg}^{\PQt\PQb}$
are defined like the $\sigma_{\MyggH}$ terms in Eq.~(\ref{eq:CgNLOQCD}).
The $\Gamma_{\Pg\Pg}^{ii}$ correspond to the partial widths that are obtained for $\Cc_i=1$ and all other $\Cc_j=0, j\neq i$.
The cross-term $\Gamma_{\Pg\Pg}^{\PQt\PQb}$ can then be derived
by calculating the SM partial width by setting $\Cc_{\PQb}=\Cc_{\PQt}=1$ and
subtracting $\Gamma_{\Pg\Pg}^{\PQt\PQt}$ and $\Gamma_{\Pg\Pg}^{\PQb\PQb}$ from it. 

\paragraph*{Effective treatment}
In the general case, without the assumptions above, possible non-zero contributions from additional
particles in the loop have to be taken into account and $\Cc_{\Pg}^2$ is then treated as an
effective coupling scale factor parameter in the fit:
% \begin{eqnarray}
%  \frac{\sigma_{\MyggH}}{\sigma_{\MyggH}^\text{SM}} &=& \Cc_{\Pg}^2
% \end{eqnarray}
$\sigma_{\MyggH}/\sigma_{\MyggH}^\text{SM} = \Cc_{\Pg}^2$.
The effective scale factor for the partial gluon width $\Gamma_{\Pg\Pg}$ should behave in a very similar way,
so in this case the same effective scale factor $\Cc_{\Pg}$ is used:
% \begin{eqnarray}
%  \frac{\Gamma_{\Pg\Pg}}{\Gamma_{\Pg\Pg}^\text{SM}}  &=& \Cc_{\Pg}^2
% \end{eqnarray}
$\Gamma_{\Pg\Pg}/\Gamma_{\Pg\Pg}^\text{SM} = \Cc_{\Pg}^2$.
As the contribution of $\Gamma_{\Pg\Pg}$ to the total width is <10\% in the SM,
this assumption is believed to have no measurable impact.

\subsubsection{Scaling of the $\PH\to\PGg\PGg$ partial decay width}
\label{sec:C_gamma}
Like in the previous section,
$\Cc_{\PGg}^2$ refers to the scale factor for the loop-induced $\PH\to\PGg\PGg$ decay.
Also for the $\PH\to\PGg\PGg$ decay NLO QCD corrections exist and are implemented in \HDECAY.
This allows to treat the scale factor for the $\PGg\PGg$ partial width
as a second order polynomial in $\Cc_{\PQb}$, $\Cc_{\PQt}$, $\Cc_{\PGt}$, and $\Cc_{\PW}$:
\begin{eqnarray}
\label{eq:CgammaNLOQCD}
\Cc_{\PGg}^2(\Cc_{\PQb}, \Cc_{\PQt}, \Cc_{\PGt}, \Cc_{\PW}, \mH) &=& \frac{\sum_{i,j}\Cc_i \Cc_j\cdot\Gamma_{\PGg\PGg}^{i j}(\mH)}{\sum_{i,j}\Gamma_{\PGg\PGg}^{ij}(\mH)}
\end{eqnarray}
where the pairs $(i,j)$ are $\PQb\PQb,\PQt\PQt,\PGt\PGt,\PW\PW,\PQb\PQt,\PQb\PGt,\PQb\PW,\PQt\PGt,\PQt\PW,\PGt\PW$.
The $\Gamma_{\PGg\PGg}^{ii}$ correspond to the partial widths that are obtained for $\Cc_i=1$ and all other $\Cc_j=0, (j\neq i)$.
The cross-terms $\Gamma_{\PGg\PGg}^{ij}, (i\neq j)$ can then be derived by calculating the partial width by setting $\Cc_i=\Cc_j=1$
and all other $\Cc_l=0, (l\neq i,j)$, and subtracting $\Gamma_{\PGg\PGg}^{ii}$ and $\Gamma_{\PGg\PGg}^{jj}$ from them. 

\paragraph*{Effective treatment}
In the general case, without the assumption above, possible non-zero contributions from additional
particles in the loop have to be taken into account
and $\Cc_{\PGg}^2$ is then treated as an effective coupling parameter in the fit.

\subsubsection{Scaling of the $\PH\to \PZ\PGg$ decay vertex}
\label{sec:C_Zgamma}
Like in the previous sections, $\Cc_{(\PZ\PGg)}^2$ refers to
the scale factor for the loop-induced $\PH\to \PZ\PGg$ decay.
This allows to treat the scale factor for the $\PZ\PGg$ partial width as a
second order polynomial in $\Cc_{\PQb}$, $\Cc_{\PQt}$, $\Cc_{\PGt}$, and $\Cc_{\PW}$:
\begin{eqnarray}
\label{eq:CZgammaNLOQCD}
\Cc_{(\PZ\PGg)}^2(\Cc_{\PQb}, \Cc_{\PQt}, \Cc_{\PGt}, \Cc_{\PW}, \mH) &=& \frac{\sum_{i,j}\Cc_i \Cc_j\cdot\Gamma_{\PZ\PGg}^{i j}(\mH)}{\sum_{i,j}\Gamma_{\PZ\PGg}^{i j}(\mH)}
\end{eqnarray}
where the pairs $(i,j)$ are $\PQb\PQb,\PQt\PQt,\PGt\PGt,\PW\PW,\PQb\PQt,\PQb\PGt,\PQb\PW,\PQt\PGt,\PQt\PW,\PGt\PW$.
The $\Gamma_{\PZ\PGg}^{ij}$ are calculated in the same way as for Eq.~(\ref{eq:CgammaNLOQCD}).
NLO QCD corrections have been computed and found to be very small \cite{Spira:1991tj}, and thus ignored here.

\paragraph*{Effective treatment}
In the general case, without the assumption above, possible non-zero contributions from additional
particles in the loop have to be taken into account
and $\Cc_{(\PZ\PGg)}^2$ is then treated as an effective coupling parameter in the fit.

\subsubsection{Scaling of the total width}
The total width $\Gamma_{\PH}$ is the sum of all Higgs partial decay widths.
Under the assumption that no additional BSM Higgs decay modes
(into either invisible or undetectable final states)
contribute to the total width,
$\Gamma_{\PH}$ is expressed as the sum of the scaled partial Higgs decay widths to SM particles,
which combine to a total scale factor $\Cc_{\PH}^2$ compared to the SM total width $\Gamma_{\PH}^\text{SM}$:
\begin{eqnarray}
  \label{eq:CH2_def}
  \Cc_{\PH}^2(\Cc_i,\mH) &=& \sum\limits_{\begin{array}{r}j=\PW\PW^{(*)},\PZ\PZ^{(*)},\PQb\PAQb,\PGtm\PGtp,\\\PGg\PGg,\PZ\PGg,\Pg\Pg,\PQt\PAQt,\PQc\PAQc,\PQs\PAQs,\PGmm\PGmp\end{array}} \frac{ \Gamma_j(\Cc_i,\mH)}{\Gamma_{\PH}^\text{SM}(\mH)}
\end{eqnarray}

\paragraph*{Effective treatment}
In the general case, additional Higgs decay modes to BSM particles cannot be excluded
and the total width scale factor $\Cc_{\PH}^2$ is treated as free parameter.

The total width $\Gamma_{\PH}$ for a light Higgs with $\mH\sim 125\UGeV$
is not expected to be directly observable at the LHC,
as the SM expectation is $\Gamma_{\PH}\sim 4\UMeV$,
several orders of magnitude smaller than
the experimental mass resolution.
There is no indication from the results observed so far that 
the natural width is broadened by new physics effects 
to such an extent that it could be directly observable.
Furthermore, as all LHC Higgs channels rely on the identification of Higgs decay products,
there is no way of measuring the total Higgs width indirectly within a coupling fit without using assumptions. 
This can be illustrated by assuming that all cross sections and partial widths are increased by a common factor
$\Cc_i^2=r>1$.
If simultaneously the Higgs total width is increased by the square of the same factor $\Cc_{\PH}^2=r^2$
(for example by postulating some BSM decay mode) the experimental visible signatures
in all Higgs channels would be indistinguishable from the SM.

Hence without further assumptions only ratios of scale factors $\Cc_i$ can be measured at the LHC,
where at least one of the ratios needs to include the total width scale factor $\Cc_{\PH}^2$.
Such a definition of ratios absorbs two degrees of freedom
(e.g.\ a common scale factor to all couplings and a scale factor to the total width)
into one ratio that can be measured at the LHC.
In order to go beyond the measurement of ratios of coupling scale factors to the
determination of absolute coupling scale factors $\Cc_i$ additional
assumptions are necessary to remove one degree of freedom.
Possible assumptions are:
\begin{itemize}
 \item No new physics in Higgs decay modes (Eq.~(\ref{eq:CH2_def})).
 \item $\Cc_{\PW}\le 1$, $\Cc_{\PZ}\le 1$. If one combines this assumption with the fact that all Higgs partial decay widths are
 positive definite and the total width is bigger than the sum of all (known) partial decay widths,
 this is sufficient to give a lower and upper bound on all $\Cc_i$ and also
 determine a possible branching ratio $\BRinv$ into final states invisible or undetectable at the LHC.
 This is best illustrated with the $\PV\PH(\PH\to\PV\PV)$ process: 
 \begin{eqnarray}
  \sigma_{\PV\PH}\cdot\text{BR}(\PH\to \PV\PV) &=& \frac{\Cc_{\PV}^2 \cdot \sigma_{\PV\PH}^\text{SM} \,\,\cdot\,\, \Cc_{\PV}^2 \cdot \Gamma^\text{SM}_{\PV}}{\Gamma_{\PH}}\nonumber\\
  \text{and\hspace{4cm}}\Gamma_{\PH} &>& \Cc_{\PV}^2 \cdot \Gamma_{\PV}^\text{SM}\label{eq:GammaH_sum_inequation}\\
  \text{give combined:\hspace{1cm}}\sigma_{\PV\PH}\cdot\text{BR}(\PH\to VV)&<&\frac{\Cc_{\PV}^2 \cdot \sigma_{\PV\PH}^\text{SM} \,\,\cdot\,\, \Cc_{\PV}^2 \cdot \Gamma^\text{SM}_{\PV}}{\Cc_{\PV}^2 \cdot \Gamma_{\PV}^\text{SM}}\nonumber\\
  \Longrightarrow\text{\hspace{4cm}}\Cc_{\PV}^2&>&\frac{\sigma_{\PV\PH}\cdot\text{BR}(\PH\to \PV\PV)}{\sigma_{\PV\PH}^\text{SM}}
 \end{eqnarray}
 If more final states are included in Eq.~(\ref{eq:GammaH_sum_inequation}), the lower bounds become tighter
 and together with the upper limit assumptions on $\Cc_{\PW}$ and $\Cc_{\PZ}$, absolute measurements are possible.
 However, uncertainties on all $\Cc_i$ can be very large depending
 on the accuracy of the $\PQb\PAQb$ decay channels that dominate the uncertainty
 of the total width sum.
\end{itemize}

In the following benchmark parametrizations always two versions are given:
one without assumptions on the total width
and one assuming no beyond SM Higgs decay modes.

\subsection{Further assumptions}

\subsubsection{Theoretical uncertainties}
The quantitative impact of theory uncertainties in the Higgs production cross sections and decay rates
is discussed in detail in \Bref{Dittmaier:2011ti}.

Such uncertainties will directly affect the determination of the scale factors.
When one or more of the scaling factors differ from 1, the uncertainty from missing higher-order
contributions will in general be larger than what was estimated in~\Bref{Dittmaier:2011ti}.
 
In practice, the cross section predictions with their uncertainties as tabulated in~\Bref{Dittmaier:2011ti}
are used as such so that for $\Cc_i=1$ the recommended SM treatment is recovered.
Without a consistent electroweak NLO calculation for deviations from the SM,
electroweak corrections and their uncertainties for the SM prediction
($\sim 5\%$ in gluon fusion production and $\sim 2\%$ in the di-photon decay)
are naively scaled together.
In the absence of explicit calculations
this is the currently best available approach
in a search for deviations from the SM Higgs prediction.

\subsubsection{Limit of the zero-width approximation}
Concerning the zero-width approximation (ZWA), it should be noted that in the 
mass range of the narrow resonance the width of the Higgs boson of the
Standard Model (SM) is more than four orders of magnitude smaller than 
its mass.
Thus, the zero-width approximation is in principle expected to
be an excellent approximation not only for a SM-like Higgs boson below $\sim 150\UGeV$ 
but also for a wide range of BSM scenarios which are
compatible with the present data. 
However, it has been shown in \Bref{Kauer:2012hd}
that this is not always the case even in the SM.
The inclusion of off-shell contributions is essential 
to obtain an accurate Higgs signal normalization at the $1\%$ precision level. 
For $\Pg\Pg\ (\to \PH) \to \PV\PV$, $\PV= \PW,\PZ$, 
${\cal O}(10\%)$ corrections occur due to an enhanced Higgs signal 
in the region $M_{\PV\PV} > 2\,M_{\PV}$, where also sizeable 
Higgs-continuum interference occurs.
However, with the accuracy anticipated to be reached in
the 2012 data these effects play a minor role.

\subsubsection{Signal interference effects}
\label{Subsub:int}
% A source of uncertainty is related to interference effects in $\PH \to 4\,$fermion
% decay. We refer to Chapter~2 of \Bref{Dittmaier:2012vm}, where it is shown that the ratio 
% of the ZWA
% %--
% \begin{equation}
% \mbox{BR}(\PH \to \PV \PV)\,\times\,\mbox{BR}^2(V \to {\overline f} f)
% \end{equation}
% %--
% over the complete result~\cite{Prophecy4f,Bredenstein:2006rh,Bredenstein:2006ha} for $\PH \to \Pep\Pem\Pep\Pem$ or 
% $\Pep\Pem\PGmp\PGmm$ is large, due to the interference (below $\PW\PW, \PZ\PZ$ thresholds),
% about $11\%$ difference.
% 
% The experimental analyses take into account the full NLO 4~fermion partial decay width.
% The partial width of the 4~lepton final state
% (usually referred to as $\PH\to \PZ\PZ^{(*)}\to 4l$ or $\PH\to \PZ\PZ^{(*)}\to 2l2j$, depending on decay mode) is scaled with $\Cc_{\PZ}^2$.
% The partial width of the low mass 2~lepton, 2~neutrino final state
% (usually referred to as $\PH\to \PW\PW^{(*)}\to l\nu\,l\nu$, although some interference with $\PH\to \PZ^{(*)}\PZ\to ll\,\nu\nu$ exists and is taken into account) is scaled with $\Cc_{\PW}^2$.
A possible source of uncertainty is related to interference effects
in $\PH \to 4\,$fermion decay.
For a light Higgs boson the decay width
into 4 fermions should always be calculated from the complete matrix
elements and not from the approximation
%--
\begin{equation}\label{eq:4fapprox}
\text{BR}(\PH \to \PV \PV)\,\cdot\,\text{BR}^2(\PV \to \Pf\PAf)
\end{equation}
%--
This approximation, based on the ZWA for the gauge boson $\PV$, neglects both off-shell
effects and interference between diagrams where the intermediate gauge
bosons couple to different pairs of final-state fermions.
As shown in Chapter~2 of \Bref{Dittmaier:2012vm}, the interference
effects not included in Eq.~(\ref{eq:4fapprox}) amount to 10\% for the
decay $\PH \to \Pep\Pem\Pep\Pem$ for a $125\UGeV$ Higgs.
Similar interference effects of the order of 5\% are found for the
$\Pep\PGne\Pem\PAGne$ and $\PQq\PAQq\PQq\PAQq$ final states.

The experimental analyses take into account the full NLO 4-fermion
partial decay width~\cite{Prophecy4f,Bredenstein:2006rh,Bredenstein:2006ha}.
The partial width of the 4-lepton final state
(usually described as $\PH\to \PZ\PZ^{(*)}\to 4\Pl$) is scaled with $\Cc_{\PZ}^2$.
Similary, the partial width of the 2-lepton, 2-jet final state
(usually described as $\PH\to\PZ\PZ^{(*)}\to 2\Pl 2\PQq$) is scaled with $\Cc_{\PZ}^2$.
The partial width of the low mass
2-lepton, 2-neutrino final state (usually described as $\PH\to \PW\PW^{(*)}\to
\Pl\PGn\,\Pl\PGn$, although a contribution of $\PH\to \PZ^{(*)}\PZ\to \Pl\Pl\,\PGn\PGn$
exists and is taken into account) is scaled with $\Cc_{\PW}^2$. 

\subsubsection{Treatment of $\Gamma_{\PQc\PAQc}$, $\Gamma_{\PQs\PAQs}$, $\Gamma_{\PGmm\PGmp}$ and light fermion contributions to loop-induced processes}
When calculating $\Cc_{\PH}^2(\Cc_i,\mH)$ in a benchmark parametrization,
the final states $\PQc\PAQc$, $\PQs\PAQs$ and $\PGmm\PGmp$
(currently unobservable at the LHC) are tied to $\Cc_i$ scale factors which can
be determined from the data.
Based on flavour symmetry considerations, the following choices are made:
\begin{eqnarray}
   \frac{\Gamma_{\PQc\PAQc}}{\Gamma_{\PQc\PAQc}^\text{SM}(\mH)}        &=& \Cc_{\PQc}^2   = \Cc_{\PQt}^2\\
   \frac{\Gamma_{\PQs\PAQs}}{\Gamma_{\PQs\PAQs}^\text{SM}(\mH)}        &=& \Cc_{\PQs}^2   = \Cc_{\PQb}^2\\
   \frac{\Gamma_{\PGmm\PGmp}}{\Gamma_{\PGmm\PGmp}^\text{SM}(\mH)} &=& \Cc_{\PGm}^2 = \Cc_{\PGt}^2
\end{eqnarray} 
Following the rationale of~Ref.~\cite[Sec. 9]{Dittmaier:2011ti},
the widths of $\Pem\Pep$, $\PQu\PAQu$, $\PQd\PAQd$ and neutrino final states are neglected.

Through interference terms,
these light fermions also contribute to the loop-induced $\Pg\Pg\to\PH$ and $\PH\to\Pg\Pg,\PGg\PGg,\PZ\PGg$ vertices.
In these cases, the assumptions $\Cc_{\PQc}=\Cc_{\PQt}$, $\Cc_{\PQs}=\Cc_{\PQb}$ and $\Cc_{\PGm}=\Cc_{\PGt}$ are made.

\subsubsection{Approximation in associated $\PZ\PH$ production}
When scaling the associated $\PZ\PH$ production mode,
the contribution from $\Pg\Pg\to \PZ\PH$ through a top-quark loop is neglected.
This is estimated to be around 5\%
of the total associated $\PZ\PH$ production cross section~\cite[Sec.~4.3]{Dittmaier:2011ti}.

\section{Benchmark parametrizations}

In putting forward a set of benchmark parametrizations based on the framework
described in the previous section several considerations were taken into
account.
One concern is the stability of the fits which typically involve several
hundreds of nuisance parameters.
With that in mind, the benchmark parametrizations avoid quotients of parameters of interest.
Another constraint that heavily shapes the exact choice of parametrization
is consistency among the uncertainties
that can be extracted in different parametrizations.
Some coupling scale factors enter linearly in loop-induced photon and gluon vertices.
For that reason, all scale factors are defined at the same power,
leading to what could appear as an abundance of squared expressions.
Finally, the benchmark parametrizations are chosen such that some
potentially interesting physics scenarios can be
probed and the parameters of interest are chosen so that at least some are expected to be determined.
 
For every benchmark parametrization, two variations are provided:
\begin{enumerate}
\item The total width is scaled assuming that there are no invisible or undetected widths.
In this case $\Cc_{\PH}^2(\Cc_i,\mH)$ is a function of the free parameters.
\item The total width scale factor is
%absorbed into the parametrization.
treated as a free parameter.
In this case no assumption is done and there will be a parameter of the form $\Cc_{ij} = \Cc_i\cdot \Cc_j / \Cc_{\PH}$.
\end{enumerate}

The benchmark parametrizations are given in tabular form where each cell corresponds
to the scale factor to be applied to a given combination of production and decay mode.

For every benchmark parametrization, a list of the free parameters and their relation to the framework parameters is provided.
To reduce the amount of symbols in the tables, $\mH$ is omitted throughout.
In practice, $\mH$ can either be fixed to a given value
or profiled together with other nuisance parameters.

\subsection{One common scale factor}
\label{sec:C}

The simplest way to look for a deviation
from the predicted SM Higgs coupling structure is to leave the overall signal strength
as a free parameter.
This is presently done by the experiments,
with ATLAS finding $\mu=1.4\pm 0.3$ at 126.0\UGeV~\cite{HiggsObsATLAS}
and CMS finding $\mu=0.87\pm 0.23$ at 125.5\UGeV~\cite{HiggsObsCMS}.

In order to perform the same fit in the context of the coupling scale factor framework, the only difference is that 
$\mu = \Cc^2\cdot \Cc^2 / \Cc^2 = \Cc^2$, where the three terms $\Cc^2$ in the intermediate expression account
for production, decay and total width scaling, respectively (Table~\ref{tab:C}). 

\begin{table}[h]
\centering

\begin{tabular}{|r|c|c|c|c|c|}
\hline
\multicolumn{6}{|l|}{\bfseries Common scale factor}\\
\multicolumn{6}{|l|}{\footnotesize Free parameter: $\Cc (= \Cc_{\PQt} = \Cc_{\PQb} = \Cc_{\PGt} = \Cc_{\PW} = \Cc_{\PZ})$.} \\
\hline
 & $\PH\to\PGg\PGg$ & $\PH\to \PZ\PZ^{(*)}$ & $\PH\to \PW\PW^{(*)}$ & $\PH\to \PQb\PAQb$ & $\PH\to\PGtm\PGtp$\\
\hline
\MyggH       & \multicolumn{5}{c|}{\multirow{5}{*}{$\Cc^2$}}\\
\MyttH & \multicolumn{5}{c|}{} \\
VBF         & \multicolumn{5}{c|}{} \\
$\PW\PH$        & \multicolumn{5}{c|}{} \\
$\PZ\PH$        & \multicolumn{5}{c|}{} \\
\hline
\end{tabular}

\caption{The simplest possible benchmark parametrization where a single scale factor applies to all production and decay modes.}
\label{tab:C}
\end{table}

This parametrization, despite providing the highest experimental precision,
has several clear shortcomings,
such as ignoring that the role of the Higgs boson in providing the masses of 
the vector bosons is very different from the role it has in providing 
the masses of fermions.

\subsection{Scaling of vector boson and fermion couplings}
\label{sec:CVCF}

In checking whether an observed state is compatible with the SM Higgs boson,
one obvious question is whether it fulfills its expected role in EWSB
which is intimately related to the coupling to the vector bosons ($\PW,\PZ$).

Therefore, assuming that the SU(2) custodial symmetry holds, in the
simplest case two parameters can be defined, one
scaling the coupling to the vector bosons, $\Cc_{\PV} (=\Cc_{\PW}=\Cc_{\PZ})$,
and one scaling the coupling common to all fermions, $\Cc_{\Pf} (=\Cc_{\PQt} = \Cc_{\PQb} = \Cc_{\PGt})$.
Loop-induced processes are assumed to scale as expected from the SM structure.

In this parametrization, presented in Table~\ref{tab:CVCF},
the gluon vertex loop is effectively a fermion loop and only the photon vertex loop requires
a non-trivial scaling, given the contributions of the top and bottom quarks, of the $\PGt$ lepton, of the $\PW$-boson, as well as their
(destructive) interference.

\begin{table}[h]
\centering

\begin{tabular}{|r|c|c|c|c|c|}
\hline
\multicolumn{6}{|l|}{\bfseries Boson and fermion scaling assuming no invisible or undetectable widths}\\
\multicolumn{6}{|l|}{\footnotesize Free parameters: $\Cc_{\PV} (= \Cc_{\PW} = \Cc_{\PZ})$, $\Cc_{\Pf} (= \Cc_{\PQt} = \Cc_{\PQb} = \Cc_{\PGt})$.} \\
\hline
 & $\PH\to\PGg\PGg$ & $\PH\to \PZ\PZ^{(*)}$ & $\PH\to \PW\PW^{(*)}$ & $\PH\to \PQb\PAQb$ & $\PH\to\PGtm\PGtp$ \\
\hline
\MyggH       & \multirow{2}{*}{$\frac{\Cc_{\Pf}^2\cdot \Cc_{\PGg}^2(\Cc_{\Pf},\Cc_{\Pf},\Cc_{\Pf},\Cc_{\PV})}{\Cc_{\PH}^2(\Cc_i)}$} & \multicolumn{2}{c|}{\multirow{2}{*}{$\frac{\Cc_{\Pf}^2\cdot \Cc_{\PV}^2}{\Cc_{\PH}^2(\Cc_i)}$}} & \multicolumn{2}{c|}{\multirow{2}{*}{$\frac{\Cc_{\Pf}^2\cdot \Cc_{\Pf}^2}{\Cc_{\PH}^2(\Cc_i)}$}} \\
\MyttH &                                                                            & \multicolumn{2}{c|}{                                                 }      & \multicolumn{2}{c|}{                                                 } \\
\hline
VBF         & \multirow{3}{*}{$\frac{\Cc_{\PV}^2\cdot \Cc_{\PGg}^2(\Cc_{\Pf},\Cc_{\Pf},\Cc_{\Pf},\Cc_{\PV})}{\Cc_{\PH}^2(\Cc_i)}$} & \multicolumn{2}{c|}{\multirow{3}{*}{$\frac{\Cc_{\PV}^2\cdot \Cc_{\PV}^2}{\Cc_{\PH}^2(\Cc_i)}$}} & \multicolumn{2}{c|}{\multirow{3}{*}{$\frac{\Cc_{\PV}^2\cdot \Cc_{\Pf}^2}{\Cc_{\PH}^2(\Cc_i)}$}} \\
$\PW\PH$        &                                                                            & \multicolumn{2}{c|}{                                                 }      & \multicolumn{2}{c|}{                                                 } \\
$\PZ\PH$        &                                                                            & \multicolumn{2}{c|}{                                                 }      & \multicolumn{2}{c|}{                                                 } \\
\hline
\hline
\multicolumn{6}{|l|}{\bfseries Boson and fermion scaling without assumptions on the total width}\\
\multicolumn{6}{|l|}{\footnotesize Free parameters: $\Cc_{\PV\PV} (= \Cc_{\PV}\cdot \Cc_{\PV} / \Cc_{\PH})$, $\Rr_{\Pf\PV} (= \Cc_{\Pf} / \Cc_{\PV})$.} \\
\hline
 & $\PH\to\PGg\PGg$ & $\PH\to \PZ\PZ^{(*)}$ & $\PH\to \PW\PW^{(*)}$ & $\PH\to \PQb\PAQb$ & $\PH\to\PGtm\PGtp$ \\
\hline
\MyggH       & \multirow{2}{*}{$\Cc_{\PV\PV}^2 \cdot \Rr_{\Pf\PV}^2 \cdot \Cc_{\PGg}^2(\Rr_{\Pf\PV},\Rr_{\Pf\PV},\Rr_{\Pf\PV},1)$} & \multicolumn{2}{c|}{\multirow{2}{*}{$\Cc_{\PV\PV}^2 \cdot \Rr_{\Pf\PV}^2$}} & \multicolumn{2}{c|}{\multirow{2}{*}{$\Cc_{\PV\PV}^2 \cdot \Rr_{\Pf\PV}^2 \cdot \Rr_{\Pf\PV}^2$}} \\
\MyttH &                                      & \multicolumn{2}{c|}{                                    } & \multicolumn{2}{c|}{                                                  } \\
\hline
VBF         & \multirow{3}{*}{$\Cc_{\PV\PV}^2 \cdot \Cc_{\PGg}^2(\Rr_{\Pf\PV},\Rr_{\Pf\PV},\Rr_{\Pf\PV},1)$}  & \multicolumn{2}{c|}{\multirow{3}{*}{$\Cc_{\PV\PV}^2$}} & \multicolumn{2}{c|}{ \multirow{3}{*}{$\Cc_{\PV\PV}^2 \cdot \Rr_{\Pf\PV}^2$}} \\
$\PW\PH$        &                                      & \multicolumn{2}{c|}{                                    } & \multicolumn{2}{c|}{                                                  } \\
$\PZ\PH$        &                                      & \multicolumn{2}{c|}{                                    } & \multicolumn{2}{c|}{                                                  } \\
\hline
\end{tabular}

{\footnotesize $\Cc_i^2 = \Gamma_{ii} / \Gamma_{ii}^\text{SM}$}

\caption{A benchmark parametrization where custodial symmetry is assumed and vector boson couplings are scaled together ($\Cc_{\PV}$)
and fermions are assumed to scale with a single parameter ($\Cc_{\Pf}$).}
\label{tab:CVCF}
\end{table}

This parametrization, though exceptionally succinct, makes a number of assumptions,
which are expected to be object of further scrutiny with the accumulation of data at the LHC.
The assumptions naturally relate to the grouping of different individual couplings or to assuming that the loop amplitudes
are those predicted by the SM.

\subsection{Probing custodial symmetry}
\label{sec:RWZ}

One of the best motivated symmetries in case the new state is responsible for electroweak symmetry breaking
is the one that links its couplings to the $\PW$ and $\PZ$ bosons. 
Since $\mathrm{SU(2)_{\PV}}$ or custodial symmetry is an approximate symmetry of the SM (e.g. 
$\Delta\rho \neq 0$),
it is important to test whether data are compatible 
with the amount of violation allowed by the SM at NLO.  

In this parametrization,
presented in Table~\ref{tab:RWZ},
$\Rr_{\PW\PZ} (=\Cc_{\PW}/\Cc_{\PZ})$
%is the main parameter of interest.
is of particular interest for probing custodial symmetry.
Though providing interesting information,
both $\Cc_{\PZ}$ and $\Cc_{\Pf}$
can be thought of as nuisance parameters when performing this fit.
In addition to the photon vertex loop not having a trivial scaling,
in this parametrization also the individual $\PW$ and $\PZ$ boson fusion contributions
to the vector boson fusion production process need to be resolved.

\begin{sidewaystable}[p]
\centering
\begin{tabular}{|r|c|c|c|c|c|}
\hline
\multicolumn{6}{|l|}{\bfseries Probing custodial symmetry assuming no invisible or undetectable widths}\\
\multicolumn{6}{|l|}{\footnotesize Free parameters: $\Cc_{\PZ}$, $\Rr_{\PW\PZ} (= \Cc_{\PW} / \Cc_{\PZ})$, $\Cc_{\Pf} (= \Cc_{\PQt} = \Cc_{\PQb} = \Cc_{\PGt})$.} \\

\hline
 & $\PH\to\PGg\PGg$ & $\PH\to \PZ\PZ^{(*)}$ & $\PH\to \PW\PW^{(*)}$ & $\PH\to \PQb\PAQb$ & $\PH\to\PGtm\PGtp$ \\
\hline
\MyggH       & \multirow{2}{*}{$\frac{\Cc_{\Pf}^2\cdot \Cc_{\PGg}^2(\Cc_{\Pf},\Cc_{\Pf},\Cc_{\Pf},\Cc_{\PZ} \Rr_{\PW\PZ})}{\Cc_{\PH}^2(\Cc_i)}$} & \multirow{2}{*}{$\frac{\Cc_{\Pf}^2\cdot \Cc_{\PZ}^2}{\Cc_{\PH}^2(\Cc_i)}$} & \multirow{2}{*}{$\frac{\Cc_{\Pf}^2\cdot (\Cc_{\PZ} \Rr_{\PW\PZ})^2}{\Cc_{\PH}^2(\Cc_i)}$} & \multicolumn{2}{c|}{\multirow{2}{*}{$\frac{\Cc_{\Pf}^2\cdot \Cc_{\Pf}^2}{\Cc_{\PH}^2(\Cc_i)}$}} \\
\MyttH &                                                                & & & \multicolumn{2}{c|}{                               } \\
\hline
VBF        & $\frac{\Cc_\mathrm{VBF}^2(\Cc_{\PZ},\Cc_{\PZ} \Rr_{\PW\PZ}) \cdot \Cc_{\PGg}^2(\Cc_{\Pf},\Cc_{\Pf},\Cc_{\Pf},\Cc_{\PZ} \Rr_{\PW\PZ})}{\Cc_{\PH}^2(\Cc_i)}$ & $\frac{\Cc_\mathrm{VBF}^2(\Cc_{\PZ},\Cc_{\PZ} \Rr_{\PW\PZ})\cdot \Cc_{\PZ}^2}{\Cc_{\PH}^2(\Cc_i)}$ & $\frac{\Cc_\mathrm{VBF}^2(\Cc_{\PZ},\Cc_{\PZ} \Rr_{\PW\PZ})\cdot (\Cc_{\PZ} \Rr_{\PW\PZ})^2}{\Cc_{\PH}^2(\Cc_i)}$  & \multicolumn{2}{c|}{$\frac{\Cc_\mathrm{VBF}^2(\Cc_{\PZ},\Cc_{\PZ} \Rr_{\PW\PZ})\cdot \Cc_{\Pf}^2}{\Cc_{\PH}^2(\Cc_i)}$} \\
\hline
$\PW\PH$        & $\frac{(\Cc_{\PZ} \Rr_{\PW\PZ})^2\cdot \Cc_{\PGg}^2(\Cc_{\Pf},\Cc_{\Pf},\Cc_{\Pf},\Cc_{\PZ} \Rr_{\PW\PZ})}{\Cc_{\PH}^2(\Cc_i)}$ & $\frac{(\Cc_{\PZ} \Rr_{\PW\PZ})^2\cdot \Cc_{\PZ}^2}{\Cc_{\PH}^2(\Cc_i)}$ & $\frac{(\Cc_{\PZ} \Rr_{\PW\PZ})^2\cdot (\Cc_{\PZ} \Rr_{\PW\PZ})^2}{\Cc_{\PH}^2(\Cc_i)}$ & \multicolumn{2}{c|}{$\frac{(\Cc_{\PZ} \Rr_{\PW\PZ})^2\cdot \Cc_{\Pf}^2}{\Cc_{\PH}^2(\Cc_i)}$} \\
\hline
$\PZ\PH$        & $\frac{\Cc_{\PZ}^2\cdot \Cc_{\PGg}^2(\Cc_{\Pf},\Cc_{\Pf},\Cc_{\Pf},\Cc_{\PZ} \Rr_{\PW\PZ})}{\Cc_{\PH}^2(\Cc_i)}$          & $\frac{\Cc_{\PZ}^2\cdot \Cc_{\PZ}^2}{\Cc_{\PH}^2(\Cc_i)}$ & $\frac{\Cc_{\PZ}^2\cdot (\Cc_{\PZ} \Rr_{\PW\PZ})^2}{\Cc_{\PH}^2(\Cc_i)}$                   & \multicolumn{2}{c|}{$\frac{\Cc_{\PZ}^2\cdot \Cc_{\Pf}^2}{\Cc_{\PH}^2(\Cc_i)}$}\\
\hline
\hline
\multicolumn{6}{|l|}{\bfseries Probing custodial symmetry without assumptions on the total width}\\
\multicolumn{6}{|l|}{\footnotesize Free parameters: $\Cc_{\PZ\PZ} (= \Cc_{\PZ}\cdot \Cc_{\PZ} / \Cc_{\PH})$, $\Rr_{\PW\PZ} (= \Cc_{\PW} / \Cc_{\PZ})$, $\Rr_{FZ} (= \Cc_{\Pf} / \Cc_{\PZ})$.} \\
\hline
 & $\PH\to\PGg\PGg$ & $\PH\to \PZ\PZ^{(*)}$ & $\PH\to \PW\PW^{(*)}$ & $\PH\to \PQb\PAQb$ & $\PH\to\PGtm\PGtp$ \\
\hline
\MyggH       & \multirow{2}{*}{$\Cc_{\PZ\PZ}^2 \Rr_{FZ}^2\cdot \Cc_{\PGg}^2(\Rr_{FZ},\Rr_{FZ},\Rr_{FZ},\Rr_{\PW\PZ})$} & \multirow{2}{*}{$\Cc_{\PZ\PZ}^2 \Rr_{FZ}^2$} & \multirow{2}{*}{$\Cc_{\PZ\PZ}^2 \Rr_{FZ}^2\cdot \Rr_{\PW\PZ}^2$} & \multicolumn{2}{c|}{\multirow{2}{*}{$\Cc_{\PZ\PZ}^2 \Rr_{FZ}^2\cdot \Rr_{FZ}^2$}} \\
\MyttH &                                                                & & & \multicolumn{2}{c|}{                               } \\
\hline
VBF         & $\Cc_{\PZ\PZ}^2 \Cc_\mathrm{VBF}^2(1,\Rr_{\PW\PZ}^2)\cdot \Cc_{\PGg}^2(\Rr_{FZ},\Rr_{FZ},\Rr_{FZ},\Rr_{\PW\PZ})$   & $\Cc_{\PZ\PZ}^2 \Cc_\mathrm{VBF}^2(1,\Rr_{\PW\PZ}^2) $ & $\Cc_{\PZ\PZ}^2 \Cc_\mathrm{VBF}^2(1,\Rr_{\PW\PZ}^2)\cdot \Rr_{\PW\PZ}^2$ & \multicolumn{2}{c|}{$\Cc_{\PZ\PZ}^2 \Cc_\mathrm{VBF}^2(1,\Rr_{\PW\PZ}^2)\cdot \Rr_{FZ}^2$} \\
\hline
$\PW\PH$        & $\Cc_{\PZ\PZ}^2 \Rr_{\PW\PZ}^2\cdot \Cc_{\PGg}^2(\Rr_{FZ},\Rr_{FZ},\Rr_{FZ},\Rr_{\PW\PZ})$ & $\Cc_{\PZ\PZ}^2\cdot \Rr_{\PW\PZ}^2$ & $\Cc_{\PZ\PZ}^2 \Rr_{\PW\PZ}^2\cdot \Rr_{\PW\PZ}^2$ & \multicolumn{2}{c|}{$\Cc_{\PZ\PZ}^2 \Rr_{\PW\PZ}^2\cdot \Rr_{FZ}^2$} \\
\hline
$\PZ\PH$        & $\Cc_{\PZ\PZ}^2\cdot \Cc_{\PGg}^2(\Rr_{FZ},\Rr_{FZ},\Rr_{FZ},\Rr_{\PW\PZ})$          & $\Cc_{\PZ\PZ}^2$               & $\Cc_{\PZ\PZ}^2\cdot \Rr_{\PW\PZ}^2$          & \multicolumn{2}{c|}{$\Cc_{\PZ\PZ}^2\cdot \Rr_{FZ}^2$} \\
\hline
\end{tabular}

{\footnotesize $\Cc_i^2 = \Gamma_{ii} / \Gamma_{ii}^\text{SM}$}

\caption{A benchmark parametrization where custodial symmetry is probed through the $\Rr_{\PW\PZ}$ parameter.}
\label{tab:RWZ}
\end{sidewaystable}

\subsection{Probing the fermion sector}
\label{sec:Rdu}
\label{sec:Rlq}

In many extensions of the SM the Higgs bosons
couple differently to different types of fermions.

Given that the gluon-gluon fusion production process is dominated by the top-quark coupling,
and that there are two decay modes involving fermions,
one way of splitting fermions that is within experimental reach is
to consider up-type fermions (top quark) and
down-type fermions (bottom quark and tau lepton) separately.
In this parametrization, presented in Table~\ref{tab:Rdu}, the relevant parameter of interest is $\Rr_{\PQd\PQu} (=\Cc_{\PQd}/\Cc_{\PQu})$,
the ratio of the scale factors
of the couplings to down-type fermions,
$\Cc_{\PQd} = \Cc_{\PGt} (= \Cc_{\PGm}) = \Cc_{\PQb} (= \Cc_{\PQs})$, and up-type fermions, $\Cc_{\PQu} = \Cc_{\PQt} (= \Cc_{\PQc})$.

\begin{table}[p]
\centering
\begin{tabular}{|r|c|c|c|c|c|}
\hline
\multicolumn{6}{|l|}{\bfseries Probing up-type and down-type fermion symmetry assuming no invisible or undetectable widths}\\
\multicolumn{6}{|l|}{\footnotesize Free parameters: $\Cc_{\PV} (= \Cc_{\PZ} = \Cc_{\PW})$, $\Rr_{\PQd\PQu} (= \Cc_{\PQd} / \Cc_{\PQu})$, $\Cc_{\PQu} (= \Cc_{\PQt})$.} \\
\hline
 & $\PH\to\PGg\PGg$ & $\PH\to \PZ\PZ^{(*)}$ & $\PH\to \PW\PW^{(*)}$ & $\PH\to \PQb\PAQb$ & $\PH\to\PGtm\PGtp$ \\
\hline
\MyggH       & $\frac{\Cc_{\Pg}^2(\Cc_{\PQu} \Rr_{\PQd\PQu},\Cc_{\PQu})\cdot \Cc_{\PGg}^2(\Cc_{\PQu} \Rr_{\PQd\PQu},\Cc_{\PQu},\Cc_{\PQu} \Rr_{\PQd\PQu},\Cc_{\PV})}{\Cc_{\PH}^2(\Cc_i)}$ & \multicolumn{2}{c|}{$\frac{\Cc_{\Pg}^2(\Cc_{\PQu} \Rr_{\PQd\PQu},\Cc_{\PQu})\cdot \Cc_{\PV}^2}{\Cc_{\PH}^2(\Cc_i)}$} & \multicolumn{2}{c|}{$\frac{\Cc_{\Pg}^2(\Cc_{\PQu} \Rr_{\PQd\PQu},\Cc_{\PQu})\cdot (\Cc_{\PQu} \Rr_{\PQd\PQu})^2}{\Cc_{\PH}^2(\Cc_i)}$} \\
\hline
\MyttH & $\frac{\Cc_{\PQu}^2\cdot \Cc_{\PGg}^2(\Cc_{\PQu} \Rr_{\PQd\PQu},\Cc_{\PQu},\Cc_{\PQu} \Rr_{\PQd\PQu},\Cc_{\PV})}{\Cc_{\PH}^2(\Cc_i)}$                 & \multicolumn{2}{c|}{$\frac{\Cc_{\PQu}^2\cdot \Cc_{\PV}^2}{\Cc_{\PH}^2(\Cc_i)}$}                 & \multicolumn{2}{c|}{$\frac{\Cc_{\PQu}^2\cdot (\Cc_{\PQu} \Rr_{\PQd\PQu})^2}{\Cc_{\PH}^2(\Cc_i)}$}  \\
\hline
VBF         & \multirow{3}{*}{$\frac{\Cc_{\PV}^2\cdot \Cc_{\PGg}^2(\Cc_{\PQu} \Rr_{\PQd\PQu},\Cc_{\PQu},\Cc_{\PQu} \Rr_{\PQd\PQu},\Cc_{\PV})}{\Cc_{\PH}^2(\Cc_i)}$} & \multicolumn{2}{c|}{\multirow{3}{*}{$\frac{\Cc_{\PV}^2\cdot \Cc_{\PV}^2}{\Cc_{\PH}^2(\Cc_i)}$}} & \multicolumn{2}{c|}{\multirow{3}{*}{$\frac{\Cc_{\PV}^2\cdot (\Cc_{\PQu} \Rr_{\PQd\PQu})^2}{\Cc_{\PH}^2(\Cc_i)}$}} \\
$\PW\PH$        &                                                                              & \multicolumn{2}{c|}{                                                 } & \multicolumn{2}{c|}{                                          } \\
$\PZ\PH$        &                                                                              & \multicolumn{2}{c|}{                                                 } & \multicolumn{2}{c|}{                                          } \\
\hline
\hline
\multicolumn{6}{|l|}{\bfseries Probing up-type and down-type fermion symmetry without assumptions on the total width}\\
\multicolumn{6}{|l|}{\footnotesize Free parameters: $\Cc_{\PQu\PQu} (= \Cc_{\PQu}\cdot \Cc_{\PQu} / \Cc_{\PH})$, $\Rr_{\PQd\PQu} (= \Cc_{\PQd} / \Cc_{\PQu})$, $\Rr_{\PV\PQu} (= \Cc_{\PV} / \Cc_{\PQu})$.} \\
\hline
 & $\PH\to\PGg\PGg$ & $\PH\to \PZ\PZ^{(*)}$ & $\PH\to \PW\PW^{(*)}$ & $\PH\to \PQb\PAQb$ & $\PH\to\PGtm\PGtp$ \\
\hline
\MyggH       & $\Cc_{\PQu\PQu}^2 \Cc_{\Pg}^2(\Rr_{\PQd\PQu}, 1)\cdot \Cc_{\PGg}^2(\Rr_{\PQd\PQu}, 1,\Rr_{\PQd\PQu}, \Rr_{\PV\PQu})$ & \multicolumn{2}{c|}{$\Cc_{\PQu\PQu}^2 \Cc_{\Pg}^2(\Rr_{\PQd\PQu}, 1)\cdot \Rr_{\PV\PQu}^2$} & \multicolumn{2}{c|}{$\Cc_{\PQu\PQu}^2 \Cc_{\Pg}^2(\Rr_{\PQd\PQu}, 1)\cdot \Rr_{\PQd\PQu}^2$} \\
\hline
\MyttH & $\Cc_{\PQu\PQu}^2\cdot \Cc_{\PGg}^2(\Rr_{\PQd\PQu}, 1,\Rr_{\PQd\PQu}, \Rr_{\PV\PQu})$                  & \multicolumn{2}{c|}{$\Cc_{\PQu\PQu}^2\cdot \Rr_{\PV\PQu}^2$}                  & \multicolumn{2}{c|}{$\Cc_{\PQu\PQu}^2\cdot \Rr_{\PQd\PQu}^2$} \\
\hline
VBF         & \multirow{3}{*}{$\Cc_{\PQu\PQu}^2 \Rr_{\PV\PQu}^2\cdot \Cc_{\PGg}^2(\Rr_{\PQd\PQu}, 1,\Rr_{\PQd\PQu}, \Rr_{\PV\PQu})$} & \multicolumn{2}{c|}{\multirow{3}{*}{$\Cc_{\PQu\PQu}^2 \Rr_{\PV\PQu}^2\cdot \Rr_{\PV\PQu}^2$}} & \multicolumn{2}{c|}{\multirow{3}{*}{$\Cc_{\PQu\PQu}^2 \Rr_{\PV\PQu}^2\cdot \Rr_{\PQd\PQu}^2$}} \\
$\PW\PH$        &                                                                           & \multicolumn{2}{c|}{                                                  } & \multicolumn{2}{c|}{                                          } \\
$\PZ\PH$        &                                                                           & \multicolumn{2}{c|}{                                                  } & \multicolumn{2}{c|}{                                          } \\
\hline
\end{tabular}

{\footnotesize $\Cc_i^2 = \Gamma_{ii} / \Gamma_{ii}^\text{SM}$, $\Cc_{\PQd} = \Cc_{\PQb} = \Cc_{\PGt}$}
\caption{A benchmark parametrization where the up-type and down-type symmetry of fermions is probed through the $\Rr_{\PQd\PQu}$ parameter.}
\label{tab:Rdu}
\end{table}

Alternatively one can consider quarks and leptons separately.
In this parametrization, presented in Table~\ref{tab:Rlq},
the relevant parameter of interest is $\Rr_{\Pl\PQq} (=\Cc_{\Pl}/\Cc_{\PQq})$,
the ratio of the coupling scale factors to leptons,
$\Cc_{\Pl} = \Cc_{\PGt} (= \Cc_{\PGm})$, and quarks, $\Cc_{\PQq} = \Cc_{\PQt} (= \Cc_{\PQc}) = \Cc_{\PQb} (= \Cc_{\PQs})$.

\begin{table}[p]
\centering
\begin{tabular}{|r|c|c|c|c|c|}
\hline
\multicolumn{6}{|l|}{\bfseries Probing quark and lepton fermion symmetry assuming no invisible or undetectable widths}\\
\multicolumn{6}{|l|}{\footnotesize Free parameters: $\Cc_{\PV} (= \Cc_{\PZ} = \Cc_{\PW})$, $\Rr_{\Pl\PQq} (= \Cc_{\Pl} / \Cc_{\PQq})$, $\Cc_{\PQq} (= \Cc_{\PQt} = \Cc_{\PQb})$.} \\
\hline
 & $\PH\to\PGg\PGg$ & $\PH\to \PZ\PZ^{(*)}$ & $\PH\to \PW\PW^{(*)}$ & $\PH\to \PQb\PAQb$ & $\PH\to\PGtm\PGtp$ \\
\hline
\MyggH       & \multirow{2}{*}{$\frac{\Cc_{\PQq}^2\cdot \Cc_{\PGg}^2(\Cc_{\PQq},\Cc_{\PQq},\Cc_{\PQq} \Rr_{\Pl\PQq},\Cc_{\PV})}{\Cc_{\PH}^2(\Cc_i)}$} & \multicolumn{2}{c|}{\multirow{2}{*}{$\frac{\Cc_{\PQq}^2\cdot \Cc_{\PV}^2}{\Cc_{\PH}^2(\Cc_i)}$}} & \multirow{2}{*}{$\frac{\Cc_{\PQq}^2\cdot \Cc_{\PQq}^2}{\Cc_{\PH}^2(\Cc_i)}$} & \multirow{2}{*}{$\frac{\Cc_{\PQq}^2\cdot (\Cc_{\PQq} \Rr_{\Pl\PQq})^2}{\Cc_{\PH}^2(\Cc_i)}$} \\
\MyttH &                                                                       & \multicolumn{2}{c|}{                                 }                 & &  \\
\hline
VBF         & \multirow{3}{*}{$\frac{\Cc_{\PV}^2\cdot \Cc_{\PGg}^2(\Cc_{\PQq},\Cc_{\PQq},\Cc_{\PQq} \Rr_{\Pl\PQq},\Cc_{\PV})}{\Cc_{\PH}^2(\Cc_i)}$} & \multicolumn{2}{c|}{\multirow{3}{*}{$\frac{\Cc_{\PV}^2\cdot \Cc_{\PV}^2}{\Cc_{\PH}^2(\Cc_i)}$}} & \multirow{3}{*}{$\frac{\Cc_{\PV}^2\cdot \Cc_{\PQq}^2}{\Cc_{\PH}^2(\Cc_i)}$} & \multirow{3}{*}{$\frac{\Cc_{\PV}^2\cdot (\Cc_{\PQq} \Rr_{\Pl\PQq})^2}{\Cc_{\PH}^2(\Cc_i)}$} \\
$\PW\PH$        &                                                                       & \multicolumn{2}{c|}{                                                 } & & \\
$\PZ\PH$        &                                                                       & \multicolumn{2}{c|}{                                                 } & & \\
\hline
\hline
\multicolumn{6}{|l|}{\bfseries Probing quark and lepton fermion symmetry without assumptions on the total width}\\
\multicolumn{6}{|l|}{\footnotesize Free parameters: $\Cc_{\PQq\PQq} (= \Cc_{\PQq}\cdot \Cc_{\PQq} / \Cc_{\PH})$, $\Rr_{\Pl\PQq} (= \Cc_{\Pl} / \Cc_{\PQq})$, $\Rr_{\PV\PQq} (= \Cc_{\PV} / \Cc_{\PQq})$.} \\
\hline
 & $\PH\to\PGg\PGg$ & $\PH\to \PZ\PZ^{(*)}$ & $\PH\to \PW\PW^{(*)}$ & $\PH\to \PQb\PAQb$ & $\PH\to\PGtm\PGtp$ \\
\hline
\MyggH       & \multirow{2}{*}{$\Cc_{\PQq\PQq}^2\cdot \Cc_{\PGg}^2(1,1,\Rr_{\Pl\PQq},\Rr_{\PV\PQq})$} & \multicolumn{2}{c|}{\multirow{2}{*}{$\Cc_{\PQq\PQq}^2\cdot \Rr_{\PV\PQq}^2$}} & \multirow{2}{*}{$\Cc_{\PQq\PQq}^2$} & \multirow{2}{*}{$\Cc_{\PQq\PQq}^2\cdot \Rr_{\Pl\PQq}^2$} \\
\MyttH &                                                                       & \multicolumn{2}{c|}{                                 }                 & &  \\
\hline
VBF         & \multirow{3}{*}{$\Cc_{\PQq\PQq}^2 \Rr_{\PV\PQq}^2\cdot \Cc_{\PGg}^2(1,1,\Rr_{\Pl\PQq},\Rr_{\PV\PQq})$} & \multicolumn{2}{c|}{\multirow{3}{*}{$\Cc_{\PQq\PQq}^2 \Rr_{\PV\PQq}^2\cdot \Rr_{\PV\PQq}^2$}} & \multirow{3}{*}{$\Cc_{\PQq\PQq}^2\cdot \Rr_{\PV\PQq}^2$} & \multirow{3}{*}{$\Cc_{\PQq\PQq}^2 \Rr_{\PV\PQq}^2\cdot \Rr_{\Pl\PQq}^2$} \\
$\PW\PH$        &                                                                       & \multicolumn{2}{c|}{                                                 } & & \\
$\PZ\PH$        &                                                                       & \multicolumn{2}{c|}{                                                 } & & \\
\hline
\end{tabular}

{\footnotesize $\Cc_i^2 = \Gamma_{ii} / \Gamma_{ii}^\text{SM}$, $\Cc_{\Pl} = \Cc_{\PGt}$}

\caption{A benchmark parametrization where the quark and lepton symmetry of fermions is probed through the $\Rr_{\Pl\PQq}$ parameter.}
\label{tab:Rlq}
\end{table}

One further combination of top-quark, bottom-quark and tau-lepton, namely
scaling the top-quark and tau-lepton with a common parameter and the
bottom-quark with another parameter, can be envisaged and readily
parametrized based on the interim framework but is not put forward as a benchmark.

\subsection{Probing the loop structure and invisible or undetectable decays}
\label{sec:CgCgam}

New particles associated with physics beyond the SM
may influence the partial width of the gluon and/or photon vertices.

In this parametrization, presented in Table~\ref{tab:CgCgam},
each of the loop-induced vertices is represented by an effective scale factor, $\Cc_{\Pg}$ and $\Cc_{\PGg}$.

Particles not predicted by the SM may also give rise to invisible or undetectable
decays.
Invisible decays might show up as a MET signature and could potentially be measured at the LHC with dedicated analyses. 
An example of an undetectable final state would be a multi-jet signature that
cannot be separated from QCD backgrounds at the LHC and hence not detected.
With sufficient data it can be envisaged to disentangle the invisible and undetectable components.

In order to probe this possibility, instead of absorbing the total width into another parameter or leaving it free,
a different parameter is introduced, $\BRinv$.
The definition of $\BRinv$ is relative to the rescaled total width, $\Cc_{\PH}^2(\Cc_i)$,
and can thus be interpreted as the invisible or undetectable fraction of the total width. 

\begin{table}[p]
\centering
\begin{tabular}{|r|c|c|c|c|c|}
\hline
\multicolumn{6}{|l|}{\bfseries Probing loop structure assuming no invisible or undetectable widths}\\
\multicolumn{6}{|l|}{\footnotesize Free parameters: $\Cc_{\Pg}$, $\Cc_{\PGg}$.} \\
\hline
 & $\PH\to\PGg\PGg$ & $\PH\to \PZ\PZ^{(*)}$ & $\PH\to \PW\PW^{(*)}$ & $\PH\to \PQb\PAQb$ & $\PH\to\PGtm\PGtp$  \\
\hline
\MyggH       & $\frac{\Cc_{\Pg}^2\cdot \Cc_{\PGg}^2}{\Cc_{\PH}^2(\Cc_i)}$ & \multicolumn{4}{c|}{$\frac{\Cc_{\Pg}^2}{\Cc_{\PH}^2(\Cc_i)}$} \\
\hline
\MyttH & \multirow{4}{*}{$\frac{\Cc_{\PGg}^2}{\Cc_{\PH}^2(\Cc_i)}$}& \multicolumn{4}{c|}{\multirow{4}{*}{$\frac{1}{\Cc_{\PH}^2(\Cc_i)}$}    } \\
VBF         &                                                   & \multicolumn{4}{c|}{                           } \\
$\PW\PH$        &                                                   & \multicolumn{4}{c|}{                           } \\
$\PZ\PH$        &                                                   & \multicolumn{4}{c|}{                           } \\
\hline
\hline
\multicolumn{6}{|l|}{\bfseries Probing loop structure allowing for invisible or undetectable widths}\\
\multicolumn{6}{|l|}{\footnotesize Free parameters: $\Cc_{\Pg}$, $\Cc_{\PGg}$, $\BRinv$.} \\
\hline
 & $\PH\to\PGg\PGg$ & $\PH\to \PZ\PZ^{(*)}$ & $\PH\to \PW\PW^{(*)}$ & $\PH\to \PQb\PAQb$ & $\PH\to\PGtm\PGtp$  \\
\hline
\MyggH       & $\frac{\Cc_{\Pg}^2\cdot \Cc_{\PGg}^2}{\Cc_{\PH}^2(\Cc_i) /(1-\BRinv)}$ & \multicolumn{4}{c|}{$\frac{\Cc_{\Pg}^2}{\Cc_{\PH}^2(\Cc_i) /(1-\BRinv)}$} \\
\hline
\MyttH & \multirow{4}{*}{$\frac{\Cc_{\PGg}^2}{\Cc_{\PH}^2(\Cc_i) /(1-\BRinv)}$}& \multicolumn{4}{c|}{\multirow{4}{*}{$\frac{1}{\Cc_{\PH}^2(\Cc_i) /(1-\BRinv)}$}    } \\
VBF         &                                                   & \multicolumn{4}{c|}{                           } \\
$\PW\PH$        &                                                   & \multicolumn{4}{c|}{                           } \\
$\PZ\PH$        &                                                   & \multicolumn{4}{c|}{                           } \\
\hline
\end{tabular}

{\footnotesize $\Cc_i^2 = \Gamma_{ii} / \Gamma_{ii}^\text{SM}$}

\caption{A benchmark parametrization where effective vertex couplings are allowed to float through the $\Cc_{\Pg}$ and $\Cc_{\PGg}$ parameters.
Instead of absorbing $\Cc_{\PH}$, explicit allowance is made for a contribution from invisible or undetectable widths via the $\BRinv$ parameter.}
\label{tab:CgCgam}
\end{table}

One particularity of this benchmark parametrization is that it should allow
%any
theoretical predictions involving new particles to be projected into the
 $(\Cc_{\Pg},\Cc_{\PGg})$ or $(\Cc_{\Pg},\Cc_{\PGg},\BRinv)$ spaces.

It can be noted that the benchmark parametrization including
\BRinv\ can be recast in a form that allows
for an interpretation in terms of a tree-level scale factor and
the loop-induced scale factors with the following
substitutions: $\Cc_j \to \Cc_j^\prime / \Cc_\mathrm{tree}$
(with $j=\Pg,\PGg$) and $(1-\BRinv) \to \Cc_\mathrm{tree}^2$.

\subsection{A minimal parametrization without assumptions on new physics contributions}

Finally, the following parametrization gathers the most
important degrees of freedom considered before,
namely $\Cc_{\Pg}$, $\Cc_{\PGg}$, $\Cc_{\PV}$, $\Cc_{\Pf}$.
The parametrization, presented in Table~\ref{tab:minimal_noassumptions},
 is chosen such that some parameters are expected to be reasonably constrained by the LHC data in the near term,
while other parameters are not expected to be as well constrained in the same time frame. 

It should be noted that this is a parametrization which only includes trivial scale factors.

With the presently available analyses and data, $\Cc_{\Pg \PV}^2 = \Cc_{\Pg}^2\cdot \Cc_{\PV}^2 / \Cc_{\PH}^2$
seems to be a good choice for the common $\Cc_{ij}$ parameter.

\begin{table}[p]
\centering
\begin{tabular}{|r|c|c|c|c|c|}
\hline
\multicolumn{6}{|l|}{\bfseries Probing loops while allowing other couplings to float assuming no invisible or undetectable widths} \\
\multicolumn{6}{|l|}{\footnotesize Free parameters: $\Cc_{\Pg}$, $\Cc_{\PGg}$, $\Cc_{\PV} (=\Cc_{\PW}=\Cc_{\PZ})$, $\Cc_{\Pf} (=\Cc_{\PQt}=\Cc_{\PQb}=\Cc_{\PGt})$.} \\
\hline
 & $\PH\to\PGg\PGg$ & $\PH\to \PZ\PZ^{(*)}$ & $\PH\to \PW\PW^{(*)}$ & $\PH\to \PQb\PAQb$ & $\PH\to\PGtm\PGtp$ \\
\hline
\MyggH       & $\frac{\Cc_{\Pg}^2\cdot \Cc_{\PGg}^2}{\Cc_{\PH}^2(\Cc_i)}$ & \multicolumn{2}{c|}{$\frac{\Cc_{\Pg}^2\cdot \Cc_{\PV}^2}{\Cc_{\PH}^2(\Cc_i)}$} & \multicolumn{2}{c|}{$\frac{\Cc_{\Pg}^2\cdot \Cc_{\Pf}^2}{\Cc_{\PH}^2(\Cc_i)}$} \\
\hline
\MyttH & $\frac{\Cc_{\Pf}^2\cdot \Cc_{\PGg}^2}{\Cc_{\PH}^2(\Cc_i)}$ & \multicolumn{2}{c|}{$\frac{\Cc_{\Pf}^2\cdot \Cc_{\PV}^2}{\Cc_{\PH}^2(\Cc_i)}$} & \multicolumn{2}{c|}{$\frac{\Cc_{\Pf}^2\cdot \Cc_{\Pf}^2}{\Cc_{\PH}^2(\Cc_i)}$} \\
\hline
VBF         & \multirow{3}{*}{$\frac{\Cc_{\PV}^2\cdot \Cc_{\PGg}^2}{\Cc_{\PH}^2(\Cc_i)}$} & \multicolumn{2}{c|}{\multirow{3}{*}{$\frac{\Cc_{\PV}^2\cdot \Cc_{\PV}^2}{\Cc_{\PH}^2(\Cc_i)}$}} & \multicolumn{2}{c|}{\multirow{3}{*}{$\frac{\Cc_{\PV}^2\cdot \Cc_{\Pf}^2}{\Cc_{\PH}^2(\Cc_i)}$}} \\
$\PW\PH$        &                                                          & \multicolumn{2}{c|}{                                                 } & \multicolumn{2}{c|}{                                                 } \\
$\PZ\PH$        &                                                          & \multicolumn{2}{c|}{                                                 } & \multicolumn{2}{c|}{                                                 } \\
\hline
\hline
\multicolumn{6}{|l|}{\bfseries Probing loops while allowing other couplings to float allowing for invisible or undetectable widths} \\
\multicolumn{6}{|l|}{\footnotesize Free parameters: $\Cc_{\Pg\PV} (= \Cc_{\Pg}\cdot \Cc_{\PV} / \Cc_{\PH})$, $\Rr_{\PV\Pg}(=\Cc_{\PV}/\Cc_{\Pg})$, $\Rr_{\PGg\PV}(=\Cc_{\PGg} /\Cc_{\PV})$, $\Rr_{\Pf\PV}(=\Cc_{\Pf}/\Cc_{\PV})$.} \\
\hline
 & $\PH\to\PGg\PGg$ & $\PH\to \PZ\PZ^{(*)}$ & $\PH\to \PW\PW^{(*)}$ & $\PH\to \PQb\PAQb$ & $\PH\to\PGtm\PGtp$ \\
\hline
\MyggH       & $\Cc_{\Pg\PV}^2\cdot \Rr_{\PGg\PV}^2$                   & \multicolumn{2}{c|}{$\Cc_{\Pg\PV}^2$}                   & \multicolumn{2}{c|}{$\Cc_{\Pg\PV}^2\cdot \Rr_{\Pf\PV}^2$} \\
\hline
\MyttH & $\Cc_{\Pg\PV}^2 \Rr_{\PV\Pg}^2 \Rr_{\Pf\PV}^2\cdot \Rr_{\PGg\PV}^2$ & \multicolumn{2}{c|}{$\Cc_{\Pg\PV}^2 \Rr_{\PV\Pg}^2 \Rr_{\Pf\PV}^2$} & \multicolumn{2}{c|}{$\Cc_{\Pg\PV}^2 \Rr_{\PV\Pg}^2 \Rr_{\Pf\PV}^2\cdot \Rr_{\Pf\PV}^2$} \\
\hline
VBF         & \multirow{3}{*}{$\Cc_{\Pg\PV}^2 \Rr_{\PV\Pg}^2\cdot \Rr_{\PGg\PV}^2$} & \multicolumn{2}{c|}{\multirow{3}{*}{$\Cc_{\Pg\PV}^2 \Rr_{\PV\Pg}^2$}} & \multicolumn{2}{c|}{\multirow{3}{*}{$\Cc_{\Pg\PV}^2 \Rr_{\PV\Pg}^2\cdot \Rr_{\Pf\PV}^2$}} \\
$\PW\PH$        &                                                          & \multicolumn{2}{c|}{                                    } & \multicolumn{2}{c|}{                                          } \\
$\PZ\PH$        &                                                          & \multicolumn{2}{c|}{                                    } & \multicolumn{2}{c|}{                                          } \\
\hline
\end{tabular}

{\footnotesize $\Cc_i^2 = \Gamma_{ii} / \Gamma_{ii}^\text{SM}$, $\Cc_{\PV} = \Cc_{\PW} = \Cc_{\PZ}$, $\Cc_{\Pf} = \Cc_{\PQt} = \Cc_{\PQb} = \Cc_{\PGt}$}

\caption{A benchmark parametrization where effective vertex couplings are allowed to float through the $\Cc_{\Pg}$ and $\Cc_{\PGg}$ parameters
and the gauge and fermion couplings through the unified parameters $\Cc_{\PV}$ and $\Cc_{\Pf}$.}
\label{tab:minimal_noassumptions}
\end{table}

% \input{newdirections/newdirections}
% \input{conclusions/conclusions}
%\clearpage

% --------------------------------------------------
%%\bibliographystyle{cernyrep}
\bibliographystyle{atlasnote}
\bibliography{LHCHXSWG-2012-001}

\cleardoublepage
% --------------------------------------------------
\appendix
\section{General parametrization}
\label{sec:maximal}

Table~\ref{tab:maximal_noassumptions} presents the relations in a fit only with simple scale factors.
It should be noted that the number of degrees of freedom is too large to make such a fit feasible in the near future.

Several choices are possible for $\Cc_{ij}$.
With the currently available channels, $\Cc_{\Pg\PZ} = \Cc_{\Pg}\cdot \Cc_{\PZ} / \Cc_{\PH}$ seems most appropriate,
as shown in table~\ref{tab:maximal_noassumptions}.
The more appealing choices using vector boson scattering
$\Cc_{\PW\PW} = \Cc_{\PW}\cdot \Cc_{\PW} / \Cc_{\PH}$ or $\Cc_{\PZ\PZ} = \Cc_{\PZ}\cdot \Cc_{\PZ} / \Cc_{\PH}$
%will not be as good
will have lower sensitivity
until more data is accumulated.

\begin{sidewaystable}
\centering

\begin{tabular}{|r|c|c|c|c|c|}
\hline                
\multicolumn{6}{|l|}{\bfseries General parametrization allowing other couplings to float} \\
\multicolumn{6}{|l|}{\footnotesize Free parameters: $\Cc_{\Pg\PZ} (= \Cc_{\Pg}\cdot \Cc_{\PZ} / \Cc_{\PH})$, $\Rr_{\PGg\PZ} (= \Cc_{\PGg} / \Cc_{\PZ})$, $\Rr_{\PW\PZ} (= \Cc_{\PW} / \Cc_{\PZ})$, $\Rr_{\PQb\PZ} (= \Cc_{\PQb} / \Cc_{\PZ})$, $\Rr_{\PGt\PZ} (= \Cc_{\PGt} / \Cc_{\PZ})$, $\Rr_{\PZ\Pg} (= \Cc_{\PZ} / \Cc_{\Pg})$, $\Rr_{\PQt\Pg} (= \Cc_{\PQt} / \Cc_{\Pg})$.} \\
\hline
                       & \tc{red}{$\PH\to\PGg\PGg$}                                                             & $\PH\to \PZ\PZ^{(*)}$                                                                   & \tc{red}{$\PH\to \PW\PW^{(*)}$}                                                                & \tc{red}{$\PH\to \PQb\PAQb$}                                                          & \tc{red}{$\PH\to\PGtm\PGtp$}                                                              \\\hline                                                                                                                                                                                                                                                                     
\MyggH                  & $\Cc_{\Pg\PZ}^2 \hcdot \tc{blue}{1}                                   \hcdot \tc{red}{\Rr_{\PGg\PZ}^2}$ & $\Cc_{\Pg\PZ}^2 \hcdot \tc{blue}{1}                                   \hcdot \tc{red}{1}$ & $\Cc_{\Pg\PZ}^2\hcdot \tc{blue}{1}                                   \hcdot \tc{red}{\Rr_{\PW\PZ}^2}$ & $\Cc_{\Pg\PZ}^2\hcdot \tc{blue}{1}                                   \hcdot \tc{red}{\Rr_{\PQb\PZ}^2}$ & $\Cc_{\Pg\PZ}^2\hcdot \tc{blue}{1}                                   \hcdot \tc{red}{\Rr_{\PGt\PZ}^2}$ \\\hline                                                                                                                                                                                                                                                                     
\tc{blue}{\MyttH} & $\Cc_{\Pg\PZ}^2 \hcdot \tc{blue}{\Rr_{\PQt\Pg}^2}                           \hcdot \tc{red}{\Rr_{\PGg\PZ}^2}$ & $\Cc_{\Pg\PZ}^2 \hcdot \tc{blue}{\Rr_{\PQt\Pg}^2}                           \hcdot \tc{red}{1}$ & $\Cc_{\Pg\PZ}^2\hcdot \tc{blue}{\Rr_{\PQt\Pg}^2}                           \hcdot \tc{red}{\Rr_{\PW\PZ}^2}$ & $\Cc_{\Pg\PZ}^2\hcdot \tc{blue}{\Rr_{\PQt\Pg}^2}                           \hcdot \tc{red}{\Rr_{\PQb\PZ}^2}$ & $\Cc_{\Pg\PZ}^2\hcdot \tc{blue}{\Rr_{\PQt\Pg}^2}                           \hcdot \tc{red}{\Rr_{\PGt\PZ}^2}$ \\\hline                                                                                                                                                                                                                                                                     
\tc{blue}{VBF}         & $\Cc_{\Pg\PZ}^2 \hcdot \tc{blue}{\Rr_{\PZ\Pg}^2 \Cc_\mathrm{VBF}^2(1,\Rr_{\PW\PZ})} \hcdot \tc{red}{\Rr_{\PGg\PZ}^2}$ & $\Cc_{\Pg\PZ}^2 \hcdot \tc{blue}{\Rr_{\PZ\Pg}^2 \Cc_\mathrm{VBF}^2(1,\Rr_{\PW\PZ})} \hcdot \tc{red}{1}$ & $\Cc_{\Pg\PZ}^2\hcdot \tc{blue}{\Rr_{\PZ\Pg}^2 \Cc_\mathrm{VBF}^2(1,\Rr_{\PW\PZ})} \hcdot \tc{red}{\Rr_{\PW\PZ}^2}$ & $\Cc_{\Pg\PZ}^2\hcdot \tc{blue}{\Rr_{\PZ\Pg}^2 \Cc_\mathrm{VBF}^2(1,\Rr_{\PW\PZ})} \hcdot \tc{red}{\Rr_{\PQb\PZ}^2}$ & $\Cc_{\Pg\PZ}^2\hcdot \tc{blue}{\Rr_{\PZ\Pg}^2 \Cc_\mathrm{VBF}^2(1,\Rr_{\PW\PZ})} \hcdot \tc{red}{\Rr_{\PGt\PZ}^2}$ \\\hline                                         
\tc{blue}{$\PW\PH$}        & $\Cc_{\Pg\PZ}^2 \hcdot \tc{blue}{\Rr_{\PZ\Pg}^2 \Rr_{\PW\PZ}^2}                  \hcdot \tc{red}{\Rr_{\PGg\PZ}^2}$ & $\Cc_{\Pg\PZ}^2 \hcdot \tc{blue}{\Rr_{\PZ\Pg}^2 \Rr_{\PW\PZ}^2}                  \hcdot \tc{red}{1}$ & $\Cc_{\Pg\PZ}^2\hcdot \tc{blue}{\Rr_{\PZ\Pg}^2 \Rr_{\PW\PZ}^2}                  \hcdot \tc{red}{\Rr_{\PW\PZ}^2}$ & $\Cc_{\Pg\PZ}^2\hcdot \tc{blue}{\Rr_{\PZ\Pg}^2 \Rr_{\PW\PZ}^2}                  \hcdot \tc{red}{\Rr_{\PQb\PZ}^2}$ & $\Cc_{\Pg\PZ}^2\hcdot \tc{blue}{\Rr_{\PZ\Pg}^2 \Rr_{\PW\PZ}^2}                  \hcdot \tc{red}{\Rr_{\PGt\PZ}^2}$ \\\hline                                         
\tc{blue}{$\PZ\PH$}        & $\Cc_{\Pg\PZ}^2 \hcdot \tc{blue}{\Rr_{\PZ\Pg}^2}                            \hcdot \tc{red}{\Rr_{\PGg\PZ}^2}$ & $\Cc_{\Pg\PZ}^2 \hcdot \tc{blue}{\Rr_{\PZ\Pg}^2}                            \hcdot \tc{red}{1}$ & $\Cc_{\Pg\PZ}^2\hcdot \tc{blue}{\Rr_{\PZ\Pg}^2}                            \hcdot \tc{red}{\Rr_{\PW\PZ}^2}$ & $\Cc_{\Pg\PZ}^2\hcdot \tc{blue}{\Rr_{\PZ\Pg}^2}                            \hcdot \tc{red}{\Rr_{\PQb\PZ}^2}$ & $\Cc_{\Pg\PZ}^2\hcdot \tc{blue}{\Rr_{\PZ\Pg}^2}                            \hcdot \tc{red}{\Rr_{\PGt\PZ}^2}$ \\\hline
\end{tabular}

{\footnotesize $\Cc_i^2 = \Gamma_{ii} / \Gamma_{ii}^\text{SM}$}

\caption{A benchmark parametrization without further assumptions and maximum degrees of freedom.
The colors denote the common factor (black) and the factors related to the production (blue) and decay modes (red). Ones are used to denote the trivial factor.}
\label{tab:maximal_noassumptions}
\end{sidewaystable}

\section{LO SM-inspired loop parametrizations}

This appendix collects LO SM-inspired relations that could be used as scale factors of couplings involving loops. We stress that these relations are not used in the present note and are not recommended. They are added only for the sake of illustration.

\subsection*{Gluon vertex loop}

Under the assumption that the only relevant contributions to $\sigma_{\MyggH}$ and $\Gamma_{\Pg\Pg}$ are from top-quark
and bottom-quark loops, $\Cc_{\Pg}^2(\Cc_{\PQb},\Cc_{\PQt},\mH)$ is a scaling function depending on the scale factors $\Cc_{\PQb}$ and $\Cc_{\PQt}$:
\begin{eqnarray}
\label{eq:CgLO}
 \Cc_{\Pg}^2(\Cc_{\PQb}, \Cc_{\PQt},\mH) &=& \frac{|\Cc_{\PQb} A_{\PQb}(\mH)+\Cc_{\PQt} A_{\PQt}(\mH)|^2}{|A_{\PQb}(\mH)+A_{\PQt}(\mH)|^2}
\end{eqnarray}
where $A_{\PQb,\PQt}$ denotes the bottom-quark and top-quark amplitudes in the SM~\cite[Eq.~(21)]{Spira:1995rr}. 

\subsection*{Photon vertex loop}

Under the assumption that the only relevant contributions to $\Gamma_{\PGg\PGg}$ are from $\PW$-boson, top-quark,
and bottom-quark loops, $\Cc_{\PGg}^2(\Cc_{\PQb}, \Cc_{\PQt}, \Cc_{\PW},\mH)$ is a scaling function depending
on the scale factors $\Cc_{\PQb}$, $\Cc_{\PQt}$ and $\Cc_{\PW}$:
\begin{eqnarray}
\label{eq:CgammaLO}
 \Cc_{\PGg}^2(\Cc_{\PQb}, \Cc_{\PQt}, \Cc_{\PW},\mH) &=& \frac{|\Cc_{\PQb} A^\prime_{\PQb}(\mH)+\Cc_{\PQt} A^\prime_{\PQt}(\mH)+\Cc_{\PW} A^\prime_{\PW}(\mH)|^2}{|A^\prime_{\PQb}(\mH)+A^\prime_{\PQt}(\mH)+A^\prime_{\PW}(\mH)|^2}
\end{eqnarray}
where $A^\prime_{\PQb,\PQt,\PW}$ denotes the bottom-quark, top-quark, and $\PW$-boson amplitudes in the SM, including color and charge factors~\cite[Eq.~(1)]{Spira:1995rr}. 

\subsection*{$\PZ\PGg$ vertex loop}

Under the assumption that the only relevant contributions to $\Gamma_{\PZ\PGg}$ are from $\PW$-boson, top-quark,
and bottom-quark loops, $\Cc_{(\PZ\PGg)}^2(\Cc_{\PQb}, \Cc_{\PQt}, \Cc_{\PW},\mH)$ is a scaling function depending
on the scale factors $\Cc_{\PQb}$, $\Cc_{\PQt}$ and $\Cc_{\PW}$:
\begin{eqnarray}
\label{eq:CZgammaLO}
 \Cc_{(\PZ\PGg)}^2(\Cc_{\PQb}, \Cc_{\PQt}, \Cc_{\PW},\mH) &=& \frac{|\Cc_{\PQb} B_{\PQb}(\mH)+\Cc_{\PQt} B_{\PQt}(\mH)+\Cc_{\PW} B_{\PW}(\mH)|^2}{|B_{\PQb}(\mH)+B_{\PQt}(\mH)+B_{\PW}(\mH)|^2}
\end{eqnarray}
where $B_{\PQb,\PQt,\PW}$ denotes the bottom-quark, top-quark, and $\PW$-boson amplitudes in the SM~\cite[Eq.~(7)]{Spira:1991tj}. 
In the SM, $\Cc_{(\PZ\PGg)}^2 \sim \Cc_{\PW}^2$ to within 10\%.
% In Zg, At ~ 0.64 and AW ~ -12, so ignoring top incurs (0.64-12)^2 / (12)^2 ~ 0.9.

\subsection*{Treatment of $m_{\PQb}$}
Wherever the $\PQb$-quark mass, $m_{\PQb}$,
appears in the $\Cc_{\Pg}^2$ and $\Cc_{(Z\PGg)}^2$ above (Eqs.~(\ref{eq:CgLO}) and (\ref{eq:CZgammaLO}), respectively),
the pole mass $M_{\PQb}=4.49\UGeV$ is used.

Based on the results of \Bref{Spira:1995rr},
for $\Cc_{\PGg}^2$, Eq.~(\ref{eq:CgammaLO}), the running mass $m_{\PQb}(\mu), \mu=\mH/2$ is used.

\end{document}